\documentclass{article}
\usepackage[square,sort,comma,numbers,compress]{natbib}
\usepackage{xcolor}
\usepackage[english]{babel}
\usepackage{tabularx}
\usepackage{multirow}
\usepackage{booktabs}
\usepackage{fixltx2e}
\usepackage[letterpaper,top=2cm,bottom=2cm,left=3cm,right=3cm,marginparwidth=1.75cm]{geometry}
\usepackage{amsmath}

\usepackage{graphicx}
\graphicspath{{figs/}}
\usepackage{float}
\usepackage{subcaption}  
\usepackage{authblk}
\usepackage[T1]{fontenc}

\usepackage[colorlinks=true, allcolors=blue]{hyperref}

\title{Overcoming the Size Limit of First Principles Molecular Dynamics Simulations with an In-Distribution Substructure Embedding Active Learner}

\author[1,3]{Lingyu Kong}
\author[1]{Jielan Li}
\author[1]{Lixin Sun}
\author[1]{Han Yang}
\author[1]{Hongxia Hao}
\author[2]{Chi Chen}
\author[4]{Nongnuch Artrith\thanks{This work was conducted during Lingyu Kong's internship and Nongnuch Artrith's visit to Microsoft Research.}}
\author[1]{Jose Antonio Garrido Torres}
\author[$\ast$1]{Ziheng Lu}
\author[1]{Yichi Zhou\thanks{Corresponding author: \\
\hspace*{2em}Ziheng Lu, Email: zihenglu@microsoft.com \\
\hspace*{2em}Yichi Zhou, Email: yichi.zhou@microsoft.com}}

\affil[1]{Microsoft Research AI4Science}
\affil[2]{Microsoft Quantum}
\affil[3]{Department of Computer Science and Technology, Tsinghua Univeristy}
\affil[4]{Debye Institute for Nanomaterials Science, Utrecht University}

\date{} 

\begin{document}
\maketitle

\begin{abstract}
Large-scale first principles molecular dynamics are crucial for simulating complex processes in chemical, biomedical, and materials sciences. However, the unfavorable time complexity with respect to system sizes leads to prohibitive computational costs when the simulation contains over a few hundred atoms in practice. We present an In-Distribution substructure Embedding Active Learner (IDEAL) to enable efficient simulation of large complex systems with quantum accuracy by maintaining a machine learning force field (MLFF) as an accurate surrogate to the first principles methods. By extracting high-uncertainty substructures into low-uncertainty atom environments, the active learner is allowed to concentrate on and learn from small substructures of interest rather than carrying out intractable quantum chemical computations on large structures. IDEAL is benchmarked on various systems and shows sub-linear complexity, accelerating the simulation thousands of times compared with conventional active learning and millions of times compared with pure first principles simulations. To demonstrate the capability of IDEAL in practical applications, we simulated a polycrystalline lithium system composed of one million atoms and the full ammonia formation process in a Haber-Bosch reaction on a 3-nm Iridium nanoparticle catalyst on a  computing node comprising one single A100 GPU and 24 CPU cores.
\end{abstract}

\section{Introduction}
First principles molecular dynamics (MD)\cite{car1985unified} play a pivotal role in simulating fundamental properties and complex processes in material science, chemistry, catalysis, and condensed-matter physics \cite{alfe1999melting, li2022hydrogen, wang2014discovering, hegedus2008microscopic, ye2021probing, ye2022photoelectron}. However, these methods are characterized by poor time complexity with respect to system sizes, e.g., density functional theory (DFT)\cite{hohenberg1964inhomogeneous, kohn1965self} has a general time complexity of $O(N^3)$.\cite{szabo2012modern, martin2020electronic} Therefore, carrying out MD simulations based on quantum chemical computations is prohibitively expensive when the system size is large. In practice, a nanosecond process with a characteristic spatial scale involving several hundred atoms can take a few months to simulate with on several hundred CPUs, using DFT. Machine learning force fields (MLFFs)\cite{behler2007generalized,shapeev2016moment,musaelian2023learning,chen2019graph,chen2022universal,park2021accurate,schutt2017schnet,wang2023visnet,zhang2018deep,batzner20223,xie2021bayesian,liao2022equiformer} offer an effective method to accelerate first principles simulations with minimum loss of accuracy by learning a machine learning model to reproduce the mapping between the atomic structures and the potential energies, i.e., the potential energy surface (PES). A few MLFF architectures have been developed to infer energy and forces on systems with millions of atoms\cite{johansson2022micron, musaelian2023scaling, jia2020pushing}. The core principle underlying these methods is the locality assumption\cite{unke2021machine, behler2007generalized, bartok2010gaussian, thompson2015spectral, zhang2018deep, shapeev2016moment, drautz2019atomic}, which involves decomposing the total energy of the system, E, into individual, local contributions, denoted as $\text{E}=\sum \text{E}_i$. Here, $\text{E}_i$ represents the energy contributed by the $i$-th atom and is further dependent on the local environment $\rho_i$, with $\rho_i$ representing all atoms inside a fixed cutoff distance $r$ around the $i$-th atom. The MLFF learns a representation of $\rho_i$ and the mapping between $\rho_i$ and their energy contributions on a local basis. Despite the effectiveness of locality in making MLFFs transferable and efficient, it also necessitates that enough structures are included in the training set to cover the complete range of local environments expected during the simulations. If the MLFF encounters an atomic structure with a local environment that does not fall into the distribution of the training dataset, referred to as an out-of-distribution (OOD) structure, the model struggles to accurately reflect the genuine dynamics of the systems. This can result in significant errors and nonphysical outcomes. In extreme cases, the simulation fails numerically\cite{fu2022forces,vandermause2020fly,vandermause2022active}. 

In principle, the OOD issue can be ameliorated by an active learner\cite{gubaev2019accelerating,jinnouchi2020fly,kulichenko2023uncertainty,podryabinkin2017active,podryabinkin2019accelerating,schran2020committee,smith2018less,vandermause2020fly,vandermause2022active,wilson2022batch,xie2023uncertainty,zhang2019active,johansson2022micron}, which maintains an uncertainty module to assess the reliability of the MLFF at each step of the simulation. If the MLFF is deemed unreliable on an atomic configuration, an additional quantum chemical computation is invoked and new data are incorporated into the training set to update the MLFF. This ensures the credibility of the dynamics at every stage. However, when it comes to complex systems with large numbers of atoms, which are often necessary to reproduce realistic conditions (\textit{e.g.}, supported catalysts or other composite materials and multi-phase reactions), conventional active learning methods still suffer from the high time complexity of first principles calculations and cannot be directly applied. In practice, when the system size reaches several hundred or thousand atoms, a single-point first principles calculation can take several days or even weeks to complete depending on the level of theory and the computational infrastructure, which is prohibitively expensive to run long simulations even with an actively learned MLFF.\cite{kuhne2020cp2k, giannozzi2020quantum, smith2020psi4}

We propose an In-Distribution substructure Embedding Active Learner (IDEAL) to efficiently simulate complex atomic systems with a large number of atoms and to train an MLFF on-the-fly with high accuracy. \textbf{Figure 1} shows the general framework of IDEAL. During the MD simulation, only a small number of local environments exhibit high uncertainty in each frame, as shown in \textbf{Figure 1a-b}. The IDEAL algorithm identifies and crops off these under-represented local environments as shown in \textbf{Figure 1c}. Rather than directly sending them to quantum chemical computations, they are subsequently embedded into a set of atoms that closely resemble well-represented environments from the training set, as shown in \textbf{Figure 1d-e}. The resulting structures, referred to as in-distribution embedded substructures (IDESs), are then utilized for first principles calculations and MLFF updates as shown in \textbf{Figure 1f}. By employing this design, IDEAL enables direct first principles calculations on representative substructures with only a small number of atoms. More importantly, this avoids the introduction of nonphysical local environments such as dangling bonds and artificial quantum-size effect, enabling the active learner to focus on the substructure of interest. As a result, IDEAL enables orders of magnitude increase in simulation efficiency and more importantly, a much more favorable scaling of $O(N)$, as shown in \textbf{Figure 1g}, without loss of accuracy. The accuracy and efficiency of the IDEAL algorithm are benchmarked on various systems ranging from single-element bulk phases to multi-element nanostructures, with sizes spanning from hundreds to a million atoms. When applied to systems of several hundred atoms, IDEAL achieves a several-thousand-fold acceleration compared to conventional active learning methods without compromising accuracy and a million-fold speed-up compared with first principles simulations. A larger system leads to even more significant acceleration. To demonstrate IDEAL's scalability and efficiency, we employ IDEAL to simulate the dynamic processes of ammonia formation during a Haber-Bosch reaction catalyzed by a 3-nm Iridium nanoparticle in gaseous N\textsubscript{2} and H\textsubscript{2} as well as the melting of a polycrystalline lithium metal which involves 1.02 million atoms. We successfully simulated these large systems with quantum accuracy within a few days using a single A100 GPU and a 24-core-CPU node and observed the entire dynamics reaction processes in both cases. The results underscore the significance of IDEAL in enabling the simulation of complex molecular dynamics processes for the study of reaction mechanisms.

\begin{figure}[H] 
  \centering 
  \includegraphics[width=\textwidth]{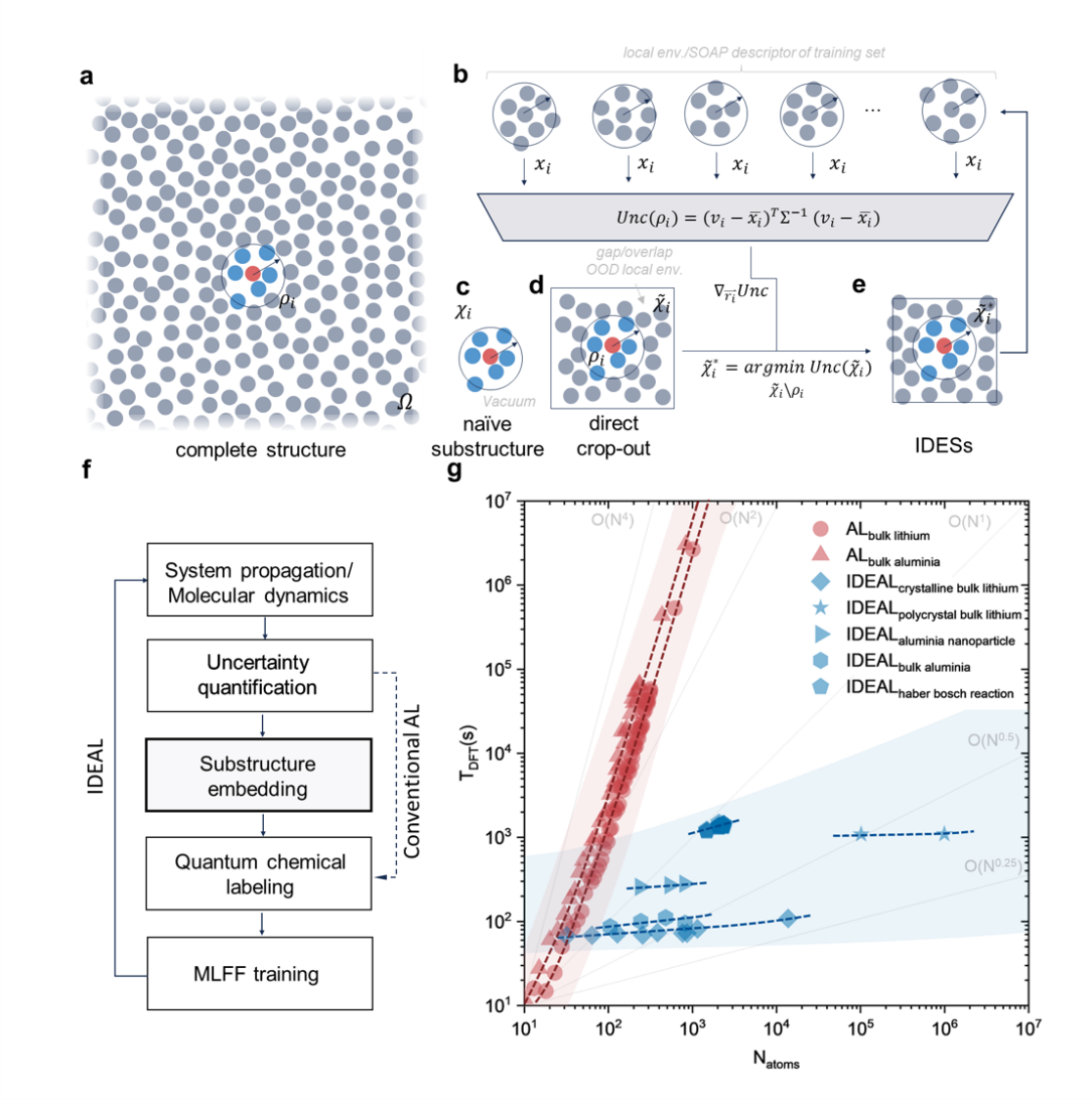} 
  \caption{\textbf{Illustration of the IDEAL algorithm. } \textbf{a} An illustrative diagram of a complex structure and one of its local environments with high uncertainty, denoted as $\rho_i$. \textbf{b} The uncertainty module estimates uncertainty for a given local environment by comparing its distribution to those in the training dataset. \textbf{c} The substructure $\chi_i$ generated by na\"ive methods. \textbf{d} The substructure $\Tilde{\chi_i}$ cropped off without minimizing the uncertainty. \textbf{e} The IDES $\Tilde{\chi_i}^*$ generated by the in-distribution substructure embedding method. \textbf{f} The overall workflow of the IDEAL algorithm and conventional active learning methods. \textbf{g} Comparison of required DFT time cost between traditional active learning methods and the IDEAL algorithm for individual high-uncertainty structures during the online MD process. The red points represent the DFT time required by traditional active learning methods for high-uncertainty Li and Al\textsubscript{2}O\textsubscript{3} structures, following approximately an $O(N^3)$ scaling. The blue data points of various shapes represent the DFT time required by the IDEAL algorithm for high-uncertainty structures with various components and sizes, following nearly a constant scaling. The dotted lines and shaded regions serve as visual guides.} 
  \label{fig:example}
\end{figure}

\section{Results}

\subsection{The IDEAL algorithm}
The overall workflow of the IDEAL algorithm closely resembles that of a conventional active learning pipeline as shown in \textbf{Figure 1f}, with the notable distinction being the substructure embedding process. The workflow is propelled by an online MD simulation. For each MD frame, we employ an uncertainty module to assess structures to detect local environments where MLFF may encounter accuracy issues. Should such environments be identified, we generate the corresponding IDESs via an uncertainty-driven in-distribution substructure embedding process and conduct DFT calculations on these substructures. The acquired substructure data is subsequently utilized for updating the MLFF. 

The in-distribution substructure embedding algorithm and the concept of IDES are central to our IDEAL algorithm. Let $\Omega$ represent the entire structure containing all atoms with periodicity, and let $\rho_i \subseteq \Omega$ denote the local environment around the $i$-th atom. Na\"ively, the substructure can be generated by cropping the selected local environment $\rho_i$ off from the entire structure and placing $\rho_i$ in vacuum to obtain structure $\chi_i$ as illustrated in \textbf{Figure 1c}. Some active learning frameworks utilize this direct method to obtain substructures for on-the-fly training of MLFFs\cite{xie2021bayesian,podryabinkin2023mlip}. Nonetheless, this direct approach leads to atoms exposed to the vacuum being situated in out-of-distribution local environments, as the atomic structures surrounding them are disrupted by the vacuum boundaries. These structures with non-physical surroundings result in noise during MLFF training, significantly impairing the model accuracy, see discussions in \textbf{Section 2.3}. To tackle the issue of out-of-distribution surroundings, we propose the in-distribution substructure embedding process. This process means embedding a high-uncertainty local environment $\rho_i$ into surroundings with low-uncertainty, \textit{i.e.}, similar to the local environments in the simulation trajectory. This involves selecting a simulation box and filling atoms around the central local environments with low overall uncertainty as a target. In principle, various methods have the potential to generate IDES, such as generative models or random structure search\cite{pickard2011ab}. 

In the current implementation of IDEAL, substructure embedding is executed through a two-step process for computational efficiency: first, substructures are extracted from the entire structure with variable cropping boundaries, and second, an uncertainty-driven optimization on the atomic coordinates is performed to obtain the IDESs. In the first step, we compute the cropping boundaries of substructures. In principle, the cropping cells should contain the local environments of the central atom of interest, \textit{i.e.}, the atoms inside the sphere as illustrated in \textbf{Figure 1a}, and be slightly larger to include some additional atoms. These additional atoms are then relaxed by minimizing the uncertainty as a target, as illustrated in \textbf{Figure 1d-e}. For simplicity, we opt for a hexahedral-shaped cropping cell and use $\Tilde{\chi_i}$ to represent the acquired substructure. Since the uncertainty changes non-smoothly and exhibits multiple local minima as the cropping boundaries move across atoms, we use a random search approach to select the best cropping cell. This also allows the choice of a simulation box with proper stoichiometry. The detailed discussions with a one-dimensional illustration are shown in \textbf{Section S1} of the Supplementary Information. In the second step, we fix the central portion $\rho_i$ and further optimize the atomic structures in the surrounding part through gradient descent with uncertainty as the target. The optimized structure is the corresponding IDES of the given $\rho_i$ and we use $\Tilde{\chi_i}^*$ to denote it for convenience. Building upon the previous description, $\Tilde{\chi_i}^*$ is given by

$$
\Tilde{\chi_i}^* = \arg\min_{\Tilde{\chi_i}\backslash \rho_i} \textup{Unc}(\Tilde{\chi_i})
$$

\noindent
where $\Tilde{\chi_i}\backslash \rho_i$ means keeping the central $\rho_i$ structure unchanged and altering the other parts of $\Tilde{\chi_i}$. This optimization result is illustrated in \textbf{Figure 1e}. The uncertainty value estimated by our uncertainty module, which will be introduced in \textbf{Section 4.2}, essentially represents the similarity between the input atomic structure and the observed structural distribution. Consequently, by minimizing uncertainty, we enhance the resemblance of the surrounding structure of $\Tilde{\chi_i}$ to the structures in the observed training distribution, which consists of structures from the previous frames of the same simulation. This process serves to reduce the noise brought by the non-physical structure, e.g., irregular bond lengths and coordination.

We evaluate the robustness and performance of IDEAL by carrying out MD simulations of bulk-phase Li, bulk-phase Al\textsubscript{2}O\textsubscript{3}, and Al\textsubscript{2}O\textsubscript{3} nanoparticles, encompassing systems with the number of atoms ranging from several hundred upto a million. The acceleration ratio achieved by IDEAL is analyzed based on these MD simulations. Lastly, we utilize the IDEAL to simulate the Haber-Bosch reaction comprising a 3 nm nanoparticle catalyst under gaseous conditions and successfully capture the entire reaction process.

\subsection{Single-Element Bulk Materials}
Lithium metal is critical in energy storage, catalysis, and a few other applications\cite{cheng2017toward, sutar2010ring}. In particular, understanding the physico-chemical behavior of lithium in its liquid state is essential to its application in, for example, interface stability of solid-state batteries\cite{han2017negating} and thermal stability of lithium-metal batteries\cite{rodrigues2017materials}. Simulating the lithium melting process typically necessitates large cell sizes containing a substantial number of atoms due to nonphysical hysteresis associated with smaller cells. Here, we utilized IDEAL to dynamically train an MLFF that can be applied across various phases and temperatures for conducting MD simulations of lithium melting. 

The simulations consist of a few consecutive runs based on IDEAL on the melting processes of lithium in crystalline phases including the hexagonal R$\bar{3}$m Li, the face-centered cubic (FCC) Li, and the body-centered cubic (BCC) Li. In each run, the initial temperature was set to 200 K. After a duration of 50 picoseconds, the system was coupled to a thermostat with a controlled temperature of 800 K, a condition that was maintained until the structure underwent complete melting. \textbf{Figure 2a-b} depicts the starting and final structure of the trajectory of a hexagonal R$\bar{3}$m Li\textsubscript{512}. During this process, as illustrated in \textbf{Figure 2c}, a small portion of the atoms experienced significant fluctuations in their atomic positions, resulting in heightened uncertainty. Consequently, IDESs were generated and were subsequently sent to DFT labeling for the MLFF update. Complete melting of lithium was observed during a relatively short amount of time due to the high temperature and large simulation cell. This is exemplified by the radial distribution function (RDF) shown in \textbf{Figure 2d} with a loss in local order. The diffusion of lithium, as tracked by the linear growth of lithium's mean square displacement (MSD), further supports the full melting of lithium as shown in \textbf{Figure 2e}. The cumulative number of DFT calculations is also shown in the figure. At 50 ps, a temperature rise prompted rapid and significant transformations of the Li structure from solid to liquid phase. Concurrently, the IDEAL algorithm detected the emergence of previously unknown local environments stemming from these structural changes, as evidenced by the swift increase in the total number of DFT calculations performed. As the MD process progressed, the system completed its phase transition. After such a transition, the newly-encountered atomic environments quickly exhausted, leading to a convergence of the accumulated number of DFT calls.

\begin{figure}[H] 
  \centering 
  \includegraphics[width=\textwidth]{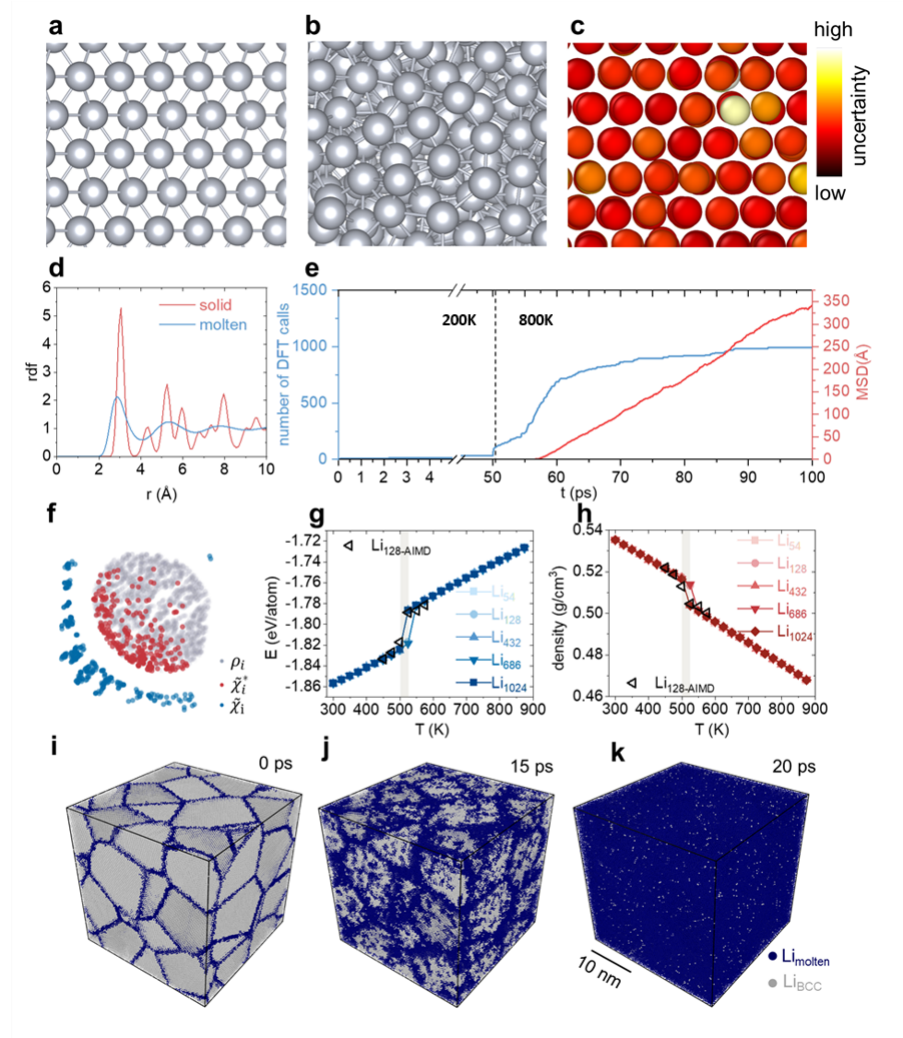} 
  \caption{\textbf{MD simulation of lithium melting with IDEAL}. \textbf{a} The initial Li\textsubscript{512} structure of the melting simulation. \textbf{b} The last frame of the Li\textsubscript{512} structure after the melting simulation. \textbf{c} A snapshot of Li\textsubscript{512} during the simulation with each atom colored by the corresponding uncertainty. \textbf{d} The RDF of lithium in solid and molten state in the simulated trajectory. \textbf{e} MSD of lithium and the accumulated number of DFT calls during a 100 ps online MD simulation. \textbf{f} The t-SNE map of SOAP features of local environments of atoms of interest. The grey points represent the SOAP features of local environments in the simulated 
  Li\textsubscript{512}. The blue and red points are the SOAP features from the surrounding parts of substructures generated by na\"ive cropping and in-distribution substructure embedding. We employ the t-SNE dimensionality reduction technique for visualizing these SOAP features in a 2D plot. \textbf{g} Average potential energy of lithium during an equilibrium MD simulation at temperatures ranging from 200K to 900K. \textbf{h} Average density of lithium during an equilibrium MD simulation at temperatures ranging from 200K to 900K. \textbf{i} A snapshot at 0ps of a lithium polycrystal containing one million atoms, simulated with the IDEAL algorithm at 600K. \textbf{j} The snapshot at 15ps. \textbf{k} The snapshot at 20ps.} 
  \label{fig:example}
\end{figure}

The accuracy of the trained MLFF was benchmarked by comparing the structures of the different phases of Li and their energetics during MD. The comparison of several ground state Li crystal structures is shown in \textbf{Section S2} in the Supplementary Information. The relative error in lattice parameters was less than 0.5\%. Evaluations on the radial distribution function, as shown in \textbf{Section S3}, also show good agreement with results from the MD simulations conducted based on first principles calculations. Furthermore, the mean absolute errors in energy (\textbf{MAE\textsubscript{E}}) and force (\textbf{MAE\textsubscript{F}}) were computed based on 600 randomly sampled frames from the MD trajectories of the Li\textsubscript{512} system and labeled through DFT computations. For both metrics the error decreased after active learning, as shown in \textbf{Table 1}. It is worth noting that such a decrease in error is not significant due to the relatively simple physics of the melting process. A more profound improvement in accuracy is showcased in the next example of Al\textsubscript{2}O\textsubscript{3}.

\begin{table*}[h]
\footnotesize
\centering
\begin{tabularx}{\textwidth}{XXX}
\toprule
{\multirow{2}{*}{\textbf{Model}}} & {\multirow{2}{*}{\textbf{MAE\textsubscript{E} (eV/atom)}}} & {\multirow{2}{*}{\textbf{MAE\textsubscript{F} (eV/Å)}}} \\
{} &  {} & {} \\
\midrule
MLFF\textsubscript{init} & 0.005 & 0.045 \\
MLFF\textsubscript{IDEAL} & 0.002 & 0.031 \\
\bottomrule
\end{tabularx}
\caption{\textbf{Performance on Li\textsubscript{512} of MLFF on-the-fly trained by IDEAL compared with initial MLFF. }}
\label{tab:label}
\end{table*}

To visualize the impact of in-distribution embedding on local environments, we generated t-SNE\cite{hinton2002stochastic} maps of the Smooth Overlap of Atomic Positions (SOAP) features\cite{bartok2013representing}  for atoms within substructures obtained through na\"ive cropping and in-distribution embedding. As illustrated in \textbf{Figure 2f}, when local environments are not embedded, the structures obtained through na\"ive cropping (depicted as blue dots) deviated significantly from those observed in physical processes (represented by grey dots). Such non-physical surroundings, such as dangling bonds, can be regarded as noise and have a detrimental impact during the machine learning model training process. In comparison, the IDESs, \textit{i.e.}, the in-distribution embedded substructures demonstrate more rational and in-distribution surrounding components (depicted as red dots). This effectively mitigated the influence of non-physical noise and ensured the accuracy of MLFF training. Detailed experimental results showcasing such an effect are discussed in the next subsection. 

We proceeded to simulate the melting of Lithium structures with varying space groups and different numbers of atoms using the IDEAL algorithm. Since the MD simulation was initiated from several different initially crystalline structures and ended in their molten states at high temperatures, the active learner should have aptly captured enough information for a complete crystal-to-liquid phase transition, allowing the MLFF obtained to generalize accurately to the simulation of lithium metal at varying temperatures and states. To substantiate this, we conducted an offline (i.e., no learning, pure inference) equilibrium MD simulation of body-centered cubic Li under ambient pressure, ranging in temperatures from 200K to 900K, employing the MLFF trained using IDEAL. Such a crystal structure is stable at ambient pressure and is observed in experiments. As depicted in \textbf{Figure 2g-h}, we observed discontinuous jumps in both density and per-atom energy profiles, signifying a pronounced melting phase transition. As the system size increased, we discerned a convergent melting point falling within the range of 500K to 525K. This finding agrees well with the independent first principles MD simulations we carried out using the same DFT setup on a 128-atom cell (see the black hollow triangles in \textbf{Figure 2g-h}) and also aligns reasonably with the experimental result of 453 K considering the overestimation inherent to equilibrium MD simulations and the utilization of the Perdew–Burke–Ernzerhof functional\cite{perdew1996generalized} in DFT computations. 

To further verify the scalability of IDEAL towards extreme system size, we further expanded the simulation to a 1.02-million-atom polycrystalline lithium system as shown in \textbf{Figure 2i-k}. For such a system, we initiate with the well-trained MLFF on lithium and carry out IDEAL-based MD simulation. The details are shown in \textbf{Section S4} in Supplementary Information. Unexpectedly, when the system becomes extremely large, the uncertainty module starts to detect OOD substructures that were not detectable in previous simulations. This further necessitates the need for online MLFF updates for large systems. As shown in \textbf{Figure 2i-k}, a clear grain boundary-initiated melting is observed, in line with the experimental grain boundary premelting in a few other materials\cite{shi2010decreasing, straumal2014grain, frolov2013structural}. It is worth noting that the current simulation is run on a single A100 GPU and a 24-core CPU node within 4 days. In contrast, computing DFT on a single frame of such a system is intractable. Therefore, the acceleration ratio of simulations on this system can only be estimated based on the scaling law, and IDEAL on this task is approximately tens of billions of times faster than conventional active learning.

\subsection{Multi-Element Bulk and Surface Systems}
Al\textsubscript{2}O\textsubscript{3} plays a pivotal role in a wide range of high-temperature applications due to its exceptional thermal stability, chemical resistance, and mechanical strength.\cite{deng2022high} In our experiments, we used bulk Al\textsubscript{2}O\textsubscript{3} and nanoparticles to validate the applicability of IDEAL to multi-element and surface systems. Cropping from such multi-element ionic systems may involve charge imbalance and these simulations are able to validate the accuracy of the IDEAL algorithm on such systems. In particular, we conducted MD simulations to replicate the melting process of bulk Al\textsubscript{2}O\textsubscript{3} with the corundum R$\bar{3}$m structure using a cell containing 810 atoms and a nanoparticle with a diameter of 2.38 nm containing 820 atoms. The MD simulation protocols were kept the same as those of bulk lithium except using a higher simulation temperature and a different initial dataset. 

Similar to the previous case, we observed compelling indications that the Al\textsubscript{2}O\textsubscript{3} crystal undergoes complete melting during the 100 ps simulation. \textbf{Figure 3a} shows the MSD of Al and O and the cumulative number of DFT calculations during this MD simulation of bulk Al\textsubscript{2}O\textsubscript{3}. The initial and last frames of the MD trajectory are shown in \textbf{Figure 3b} and \textbf{Figure 3c}, respectively. The radial distributions of atoms are presented in \textbf{Figure 3d-f}, which clearly depict the vanishing of local order after melting.

\begin{figure}[H] 
  \centering 
  \includegraphics[width=\textwidth]{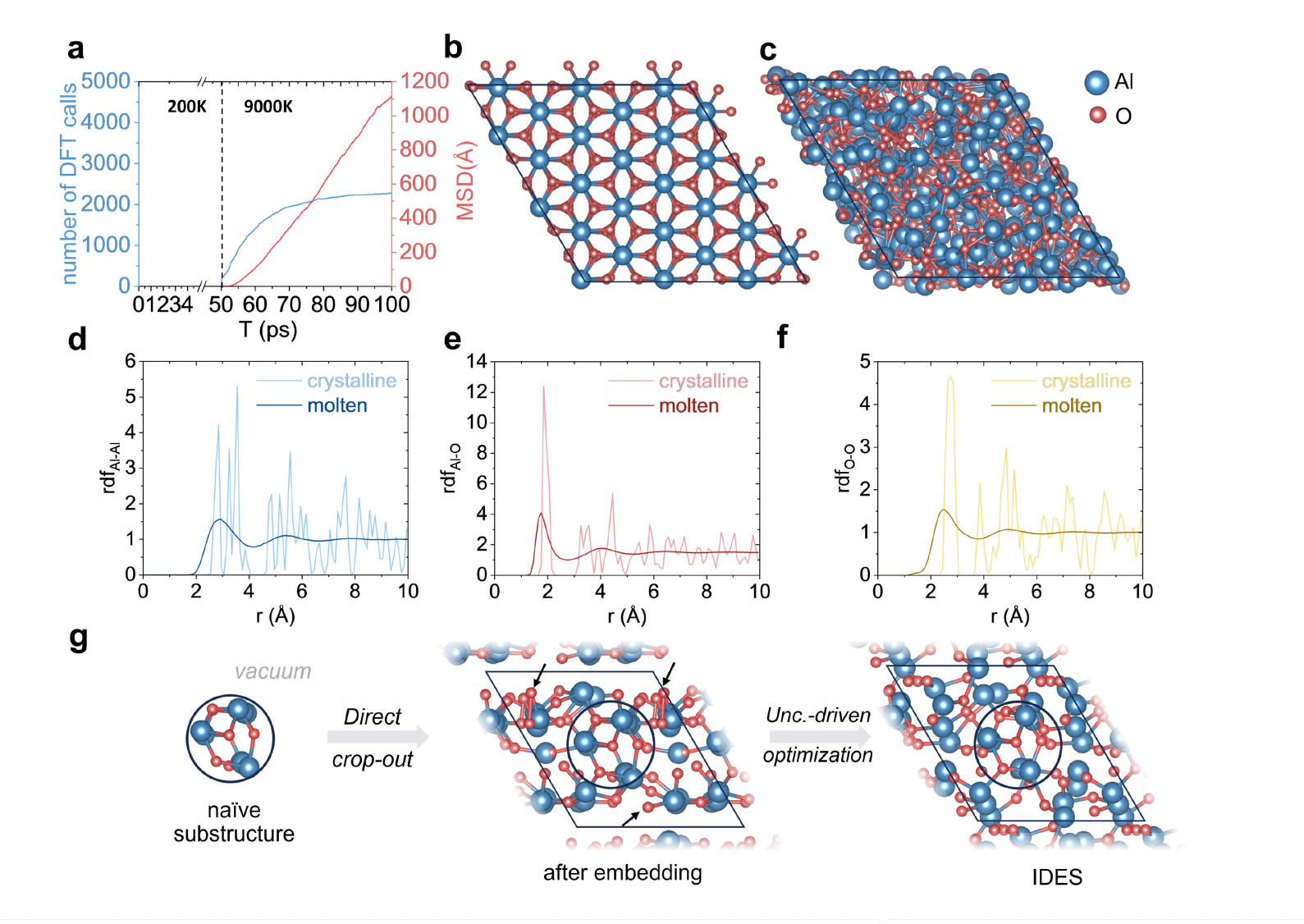} 
  \caption{\textbf{Melting simulations of crystalline Al\textsubscript{2}O\textsubscript{3}. } \textbf{a} MSD of Al and O, and the accumulated number of DFT calls during a 100 ps on-the-fly MD simulation using IDEAL. The simulation starts with a solid bulk-phase Al\textsubscript{2}O\textsubscript{3} crystal with 810 atoms under 200K and coupled to a thermostat of 9000K at $t=50ps$. \textbf{b} The initial Al\textsubscript{2}O\textsubscript{3} structure. \textbf{c} The last frame of the MD simulation. \textbf{d} The RDF between Al and Al of the initial structure and the melted structures. \textbf{e} The RDF between Al and O of the initial structure and the melted structures. \textbf{f} The RDF between O and O of the initial structure and the melted structures. \textbf{g} Substructure $\rho_i$ generated by na\"ive method, substructure $s_i$ obtained before uncertainty-driven optimization, and IDES $\Tilde{\rho_i}$ obtained after performing uncertainty-driven optimization on $s_i$.} 
  \label{fig:example}
\end{figure}

To validate the effectiveness of the active learning process, we randomly choose 600 frames from the entire MD trajectory and assess the performance of the on-the-fly learned MLFF by brute-force DFT computations on the entire structure of the corresponding frames. Similarly, we use MLFF\textsubscript{init} to represent the initial MLFF, which is trained on 2000 bulk-phase Al\textsubscript{12}O\textsubscript{18} structures, and use MLFF\textsubscript{IDEAL} for the on-the-fly learned MLFF. The results are displayed in \textbf{Table 2}. This more than 10-fold decrease in energy error substantiates the effectiveness of the active learning process. The final MLFF is benchmarked by computing ground state structure and the melting temperature of bulk Al\textsubscript{2}O\textsubscript{3} as shown in \textbf{Section S2} and \textbf{Section S5} in Supplementary Information. Following the same protocol of bulk lithium, the error in the lattice parameter of the ground state structure is negligible compared with DFT. The melting temperature is predicted to be between 2550K and 2600K, which is in reasonable agreement with the experimental value of 2345K.

An ablation study was further carried out to show how the substructure embedding affects the training accuracy and stability of the MLFF. By cropping out the substructure and carrying out the embedding with different methods, we conducted MLFF training on three datasets: 1) the na\"ive cropped-out substructures in a vacuum unit cell with dangling bonds $\chi_i$, 2) the na\"ive embedded substructures without uncertainty-driven optimization $\Tilde{\chi_i}$, and 3) the IDESs $\Tilde{\chi_i}^*$, as shown in \textbf{Figure 3g}. The MLFFs trained with different kinds of substructure data are denoted as MLFF\textsubscript{$\chi_i$}, MLFF\textsubscript{$\Tilde{\chi_i}$}, and MLFF\textsubscript{$\Tilde{\chi_i}^*$} respectively, where MLFF\textsubscript{$\Tilde{\chi_i}^*$} is also known as MLFF\textsubscript{IDEAL}. We observe an over 200-fold increase in energy error when the substructures are placed in a vacuum na\"ively and a 30-fold increase when they are not optimized for lower uncertainty as listed in \textbf{Table 2}. The results indicate that the MLFF trained with substructure data obtained from the in-distribution substructure embedding algorithm used by the IDEAL algorithm can achieve quantum accuracy, whereas the na\"ive substructure method cannot guarantee this. Moreover, it shows the in-distribution substructure embedding procedure is essential in maintaining the accuracy of the MLFF training and the robustness of on-the-fly MD simulations.

\begin{table*}[h]
\footnotesize
\centering
\begin{tabularx}{\textwidth}{XXX}
\toprule
{\multirow{2}{*}{\textbf{Model}}} & {\multirow{2}{*}{\textbf{MAE\textsubscript{E} (eV/atom)}}} & {\multirow{2}{*}{\textbf{MAE\textsubscript{F} (eV/Å)}}} \\
{} &  {} & {} \\
\midrule
MLFF\textsubscript{init} & 0.079 & 0.056 \\
MLFF\textsubscript{$\chi_i$} & 0.433 & 0.217 \\
MLFF\textsubscript{$\Tilde{\chi_i}$} & 0.068 & 0.096 \\
MLFF\textsubscript{IDEAL} (MLFF\textsubscript{$\Tilde{\chi_i}^*$}) & 0.004 & 0.045 \\
\bottomrule
\end{tabularx}
\caption{\textbf{Performance of MLFF trained on initial dataset and MLFFs actively learned on substructures $\chi_i$ that are directly cropped out and placed in vacuum, substructures $\Tilde{\chi_i}$ cropped off and placed into a periodic cell without uncertainty optimization, and IDES $\Tilde{\chi_i}^*$ that are cropped off and optimized using uncertainty as a target. }}
\label{tab:label}
\end{table*}

To further verify the capability of the current IDEAL algorithm on surface systems, we conducted a melting simulation on Al\textsubscript{2}O\textsubscript{3} nanoparticles. Similar to previous cases, quantum chemically accurate on-the-fly simulation was achieved with IDEAL. Details of this experiment are shown in \textbf{Section S6} in Supplementary Information. The result confirms that the IDEAL algorithm is capable of handling MD simulations involving surface structures and multiple elements.

\subsection{Heterogenous Catalysis}
The \textit{in silico }experiments described above were conducted to evaluate the robustness of IDEAL in single-element bulk systems, multi-element bulk systems, and surface systems. To further showcase its capability, we present the simulation results of a complex Haber-Bosch reaction catalyzed by an Iridium nanoparticle catalyst with a 3 nm diameter, conducted in a gaseous environment containing H\textsubscript{2} and N\textsubscript{2}. The Haber-Bosch reaction is an important industrial process used to convert nitrogen (N\textsubscript{2}) and hydrogen (H\textsubscript{2}) into ammonia (NH\textsubscript{3})\cite{macfarlane2020roadmap}. The structure we used for Haber-Bosch simulations comprises more than 2000 atoms as shown in \textbf{Figure 4a}. Carrying out first principles MD simulations of the Haber-Bosch reaction with such a large-scale system provides more accurate and realistic information about reaction mechanisms and catalytic performance but has been challenging due to the unfavorable scaling of DFT computations. More importantly, such a kinetic process involves rare events that are difficult to presume, necessitating on-the-fly learning.

\begin{figure}[H] 
  \centering 
  \includegraphics[width=\textwidth]{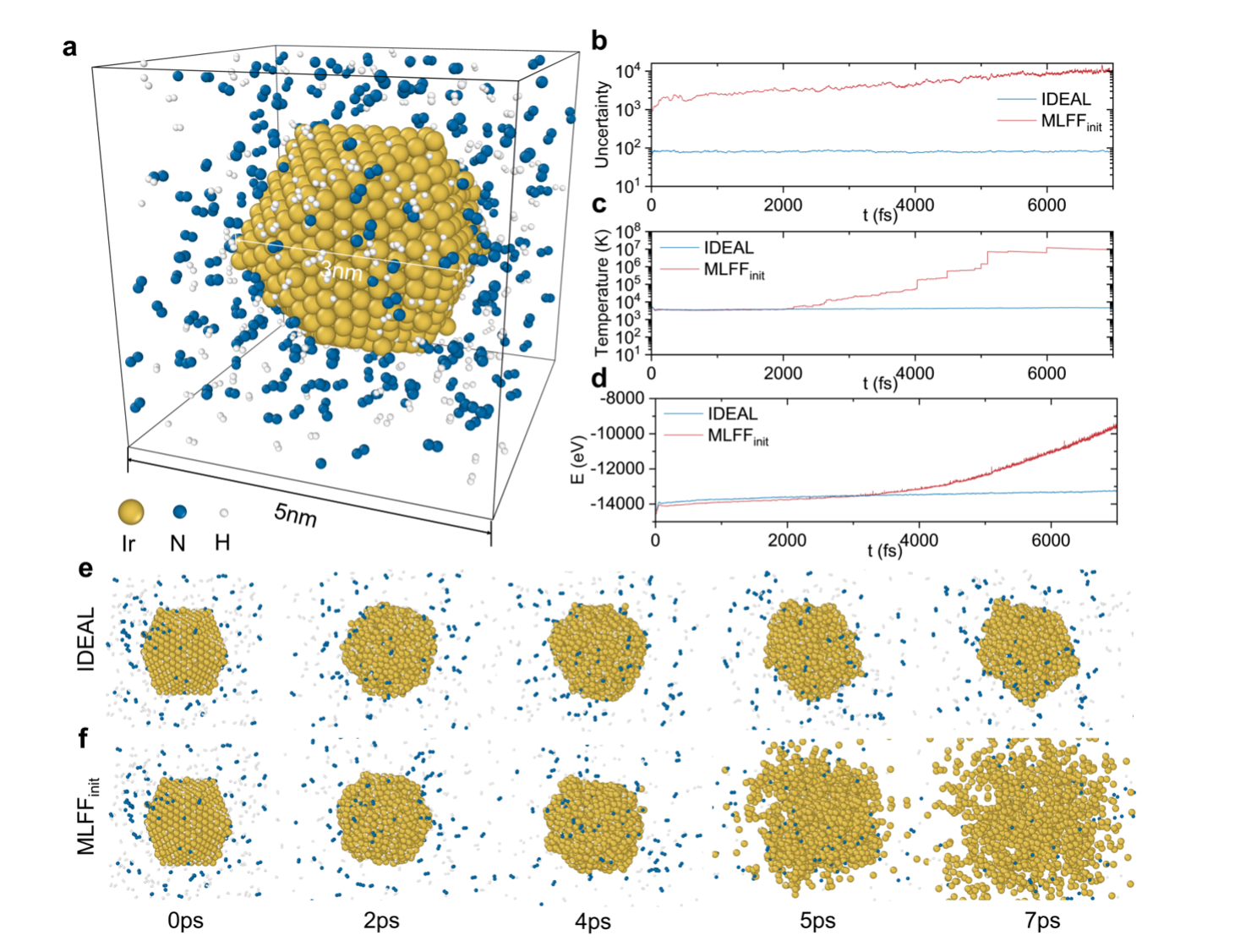}
  \caption{\textbf{Comparison between MD simulation results using IDEAL and non-actively learned MLFF. }\textbf{a} The initial configuration of the simulation cell of the Haber-Bosch reaction. \textbf{b} The uncertainty trajectories of the MD simulations based on initial MLFF and actively learned MLFF. \textbf{c} The temperature trajectories of the MD simulations based on initial MLFF and actively learned MLFF. \textbf{d} The energy trajectories of the MD simulations based on initial MLFF and actively learned MLFF. \textbf{e} The visualized trajectory of the MD simulation based on actively learned MLFF. \textbf{f} The visualized trajectory of the MD simulation based on initial MLFF. } 
  \label{fig:example}
\end{figure}

To first show the necessity of active learning, we conducted MD simulations of the Haber-Bosch reaction at a relatively high temperature of 5400K with and without the active learner. The details of the constructing the initial MLFF and the following simulation setups are discussed in \textbf{Section S7} of \textbf{Supplementary Information}. The evolution of temperature, uncertainty, and potential energy are shown in \textbf{Figure 4b-d}. Without on-the-fly learning using IDEAL, the simulation diverged with a nonphysical surge of temperature and potential energy while IDEAL kept the trajectory physical. The origin of such a nonphysical surge can be easily located by monitoring the uncertainty of the system. As shown in \textbf{Figure 4b}, the uncertainty of the system quickly grew, resulting from atomic configurations unseen by the machine learning model. This further gave rise to unreliable prediction of forces which further drove the trajectory away. In contrast, IDEAL learned from the configurations that were detected to be OOD by the uncertainty module and kept the uncertainty under a threshold. As shown in \textbf{Figure 4e-f}, the trajectory generated by the actively learned model showed reasonable melting behavior of the Iridium particle in the gaseous environment while the particle exploded in the trajectory simulated by MLFF trained solely by the predefined dataset. 

To further enhance the credibility of the simulation and train a better MLFF for more robust simulations, we further carried out several consecutive IDEAL simulations under different conditions with details discussed in \textbf{Section S7} of \textbf{Supplementary Information}. During the online simulation/training of the MLFF, several crucial steps were reproduced, including the adsorption and dissociation of H\textsubscript{2} and N\textsubscript{2} on Iridium surfaces, the formation of NH, NH\textsubscript{2}, and NH\textsubscript{3} adsorbates through the combination of N* and H*, and finally, the desorption of NH\textsubscript{3}, as shown in \textbf{Supplementary Figure 6} in \textbf{Section S7}. It is important to note that all these steps apart from N\textsubscript{2} dissociation were simulated at 1200K in a single run while the breaking of N\textsubscript{2} bonds were modeled at an elevated temperature of 7000K. 

Running these simulations enabled us to gather enough statistics for a production run of the catalytic reaction with maximum efficiency and the least cost on DFT computation of newly encountered OOD structures. The production run was carried out at 5400K by fixing the core parts of the Ir atoms while allowing the surface layer to move. Such a setup allows us to speed up the sampling process while retaining the solid nature of the catalyst. During the simulation by IDEAL, we observed all elementary steps of the catalytic formation of ammonia as discussed above. As shown in \textbf{Figure 5}, dissociation of H\textsubscript{2} happens relatively fast and was observed after only 60~fs. This indicates a high bond-breaking frequency due to the catalytic effect of Ir, which was also exemplified by the statistical evolution of chemical bonds as shown in \textbf{Supplementary Figure 7}. In contrast, the dissociation of N\textsubscript{2} became the bottleneck and was only observed a few times. The combination events of H* with N* occurred via the diffusion of H-adatoms on the Iridium particle surface. Finally, after 15~ps, the ammonia molecule detached from the Iridium particle surface, completing the entire catalytic cycle. 

\begin{figure}[H] 
  \centering 
  \includegraphics[width=\textwidth]{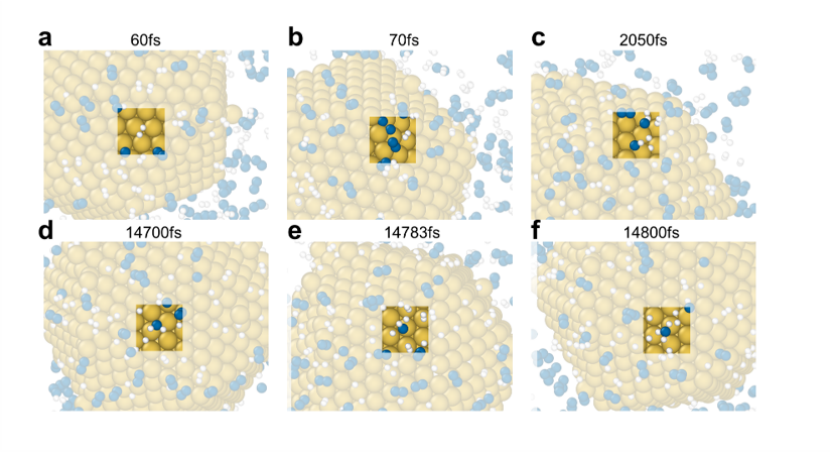}
  \caption{\textbf{Complete Haber-Bosch reaction under 5400K simulated using the MLFF trained with collected in-distribution embedded substructure data. } } 
  \label{fig:example}
\end{figure}

To assess the precision of our simulation, we further benchmarked the final MLFF using adsorption energies of N\textsubscript{2}, N, H\textsubscript{2}, H, NH, NH\textsubscript{2}, and NH\textsubscript{3} on Iridium surfaces, as presented in \textbf{Table 3}. Additional details regarding our adsorption energy estimation method can be found in \textbf{Section S}7 of the Supplementary Information. For the majority of entries in \textbf{Table 3}, the mean absolute errors (denoted as MAE) between the adsorption energies estimated by the MLFF (denoted as $\hat E_{\text{ads}}$) and those obtained through DFT calculations (denoted as $E_{\text{ads}}$) are within 70 meV. Since adsorption and desorption of these adsorbates are critical intermediate reaction steps, this indicates reasonable kinetic characteristics are reflected by the simulation. Meanwhile, in the simulation experiments of Haber-Bosch reactions, the simulated system includes Iridium nanoparticles and gas molecules, totaling about 1,600 to 3,000 atoms. Conducting first principles MD simulations on systems of this scale becomes impractical without substantial architectural enhancements due to the prohibitive time and memory requirements, while our IDEAL algorithm can complete a 100 ps simulation in a few days. Collectively, these experimental findings demonstrate the capability of the IDEAL algorithm to swiftly and accurately simulate the Haber-Bosch reaction.

\begin{table}[H]
\footnotesize
\centering
\begin{tabularx}{\textwidth}{XXXX}
\toprule
{\multirow{2}{*}{\textbf{Molecule}}} & {\multirow{2}{*}{$E_{\text{ads}}$ (eV)}} & {\multirow{2}{*}{$\hat E_{\text{ads}}$(eV)}} & {\multirow{2}{*}{MAE(eV)}} \\
{} & {} & {} {} \\
\midrule
N & -9.623 & -9.619 & 0.004 \\
N\textsubscript{2} & -1.747 & -1.762 & 0.015 \\
H & -2.992 & -2.995 & 0.003 \\
H\textsubscript{2} & -1.388 & -1.454 & 0.066 \\
NH & -6.802 & -6.892 & 0.090 \\
NH\textsubscript{2} & -5.886 & -5.913 & 0.027 \\
NH\textsubscript{3} & -2.790 & -2.698 & 0.092 \\
\bottomrule
\end{tabularx}
\caption{\textbf{Adsorption energies computed by DFT, MLFF, and their difference. }}
\label{tab:label}
\end{table}

\subsection{Scaling and Acceleration}
Quantum chemical computations are the bottlenecks of first principles MD simulations due to the unfavorable scaling with respect to system sizes. It has been elaborated that the time cost $t_{\text{DFT}}$ needed for typical DFT calculations scales cubicly with the number of valence electrons $O(N_e^3)$ and has roughly the same complexity with respect to the number of atoms $N$\cite{goedecker1999linear}. For an MD simulation of $K$ steps, the total DFT computational cost is $\text{t}_{\text{FPMD}}= O(KN^3)$. When conventional active learning is incorporated, the same scaling persists for those frames sampled, but the prefactor is reduced to a system-dependent $K^\prime$, where $K^\prime$ denotes the number of frames sent to quantum chemical computations. Thus, the computational cost of conventional active learning scales $O(K'N^3)$, resulting in an acceleration ratio of ${K}/{K^\prime}$. Despite the significant speedup, the cubic-scaled computational bottleneck remains unsolved due to the DFT calculations of the entire structure.

In contrast, IDEAL conducts DFT calculations on small substructures with a bounded number of atoms. This leads to significantly lowered costs, especially when the system size increases. Let $\bar L$ denote the average number of substructures to compute per frame from the $K'$  steps, the total DFT time for the IDEAL algorithm is $\text{T}_{\text{IDEAL}}=K' t_{sub} \bar L$, where $t_{sub}$ is the DFT time for each substructure. Assuming that the number of atoms in substructures is bounded by $m$, $t_{sub}$ scales cubicly with $m$. Therefore, the overall DFT computational complexity of IDEAL is $\text{T}_{\text{IDEAL}}= O(m^3 \bar L K')$. Since $\bar L$ is bounded by 0 and N, in theory, $\text{T}_{\text{IDEAL}}$ scales constantly to linearly with respect to system size $N$. In practice, the acceleration ratio depends on the value of $\bar L$ as well as the specific DFT parameters employed, both of which can vary on a case-by-case basis. We conduct multiple IDEAL-based MD simulations with various structure sizes and visualize the distribution of the value of $L^{(i)}$ which denotes the number of substructures extracted from the $i$-th entire structure in these experiments. The results are shown in \textbf{Section S8} of Supplementary Information. The results show that the value of $L^{(i)}$ approaches $O(1)$ when the simulation is long enough. 

To ensure the reliability of our analytical results, we conducted online MD simulations on structures of varying sizes, encompassing different systems including Lithium, Al\textsubscript{2}O\textsubscript{3}, and the Ir-N-H system. As shown in \textbf{Figure 1g}, the DFT compute time on the frames with actively learned samples is significantly lower than those compared with pure DFT on entire complex structures. In fact, when encountering OOD substructures, the time complexity of our method is found to be sub-linear, in agreement with previous analytical results. This is in sharp contrast with the near $O(N^3)$ of traditional active learning and first principles MD. Such scaling leads to a speedup of several hundred thousand times, especially when the number of atoms reaches several thousand. To complement \textbf{Figure 1g}, we list two detailed DFT time costs on Li\textsubscript{512} and Al\textsubscript{324}O\textsubscript{486} experiments and the corresponding acceleration ratios $R_{\text{IDEAL-FPMD}}$ (IDEAL versus first principles MD) and $R_{\text{IDEAL-AL}}$ (IDEAL versus conventional active learning) in \textbf{Table 5}. Here, $t_{sub}$ and $t_{entire}$ denote the average DFT time for IDESs and the entire simulated structures, respectively. In both cases, the acceleration ratio is over one thousand times compared with conventional active learning and several hundred thousand times compared with first principles MD. With larger system sizes, the value goes up even further. In fact, for system sizes larger than a few tens of thousands of atoms, conventional active learning, and first principles MD are computationally intractable on common infrastructures, whereas the IDEAL is still feasible to run. Therefore it is relatively difficult to directly measure the acceleration ratio in these cases.

\begin{table}[H]  
\centering  
\caption{\textbf{Acceleration on DFT time achieved by IDEAL in Li\textsubscript{512} and Al\textsubscript{324}O\textsubscript{486} experiments. }}  
\label{tab:acceleration}  
\begin{tabularx}{\textwidth}{*{8}{>{\centering\arraybackslash}X}}  
\toprule  
{} & {$t_{\text{sub}}/s$} & {$t_{\text{entire}}/s$} & {$K$} & {$|K'|$} & {$\bar{L}$} & {$R_{\text{IDEAL-FPMD}}$} & {$R_{\text{IDEAL-AL}}$} \\  
\midrule  
Li\textsubscript{512} & 59 & 290,828 & 100,000 & 901 & 1.104 & 495,904 & 4,463 \\  
Al\textsubscript{324}O\textsubscript{486} & 236 & 2,777,675 & 100,000 & 1,922 & 1.185 & 516,673 & 9,930 \\  
\bottomrule  
\end{tabularx}  
\end{table}

\section{Discussion}
Fast first principles simulations on large complex systems play a crucial role in providing atomic-level insights into various fields, including materials science, chemical engineering, and condensed matter physics. However, the applicability of direct first principles simulations and conventional active learning methods is limited due to their unfavorable scaling with system sizes, prohibiting their use for large realistic systems. 

In this work, we introduce the IDEAL algorithm, a solution for conducting large-scale first-principles MD simulations of complex systems that combines efficiency with unwavering accuracy. The carefully designed in-distribution substructure embedding technique seamlessly integrates high-uncertainty local environments into the substructure, effectively reducing the noise originating from non-physical local structures during MLFF training. This incorporation of substructures eliminates the need for direct quantum chemical computations on large and complex systems, resulting in significant acceleration and favorable sub-linear scaling concerning the number of atoms involved, compared to the $O(N^3)$ scaling of conventional active learning and first principles MD. By distributing individual quantum chemical tasks among multiple computational nodes, parallelization further enhances the computational capabilities of modern computing infrastructures and maximizes the utilization of available resources. The versatility of the IDEAL algorithm in handling diverse structural scenarios is demonstrated through experimental studies, including various melting processes and simulations of the Haber-Bosch reaction with system sizes of up to a million atoms.

Despite these advancements, there are some limitations and possible areas of improvement for the IDEAL algorithm. The current implementation of the algorithm is primarily focused on short-range interactions like chemical bonds or metallic interactions and may not be suitable for systems dominated by long-range interactions, such as van der Waals or Coulomb forces. As a result, while the general idea is applicable to systems like polymers or proteins, the algorithm's primary use should concentrate on local chemical reactions rather than phenomena dominated by long-range interactions, such as folding states of soft matter. To enable efficient simulation of organic compounds, the dangling bonds after cropping need to be saturated with hydrogen followed by the current uncertainty-driven optimization. Additionally, the IDEAL algorithm could potentially be interfaced with other MLFF architectures and uncertainty quantification methods to further improve its speed, scalability, and accuracy. For instance, it could be integrated with other architectures other than graph-based neural networks, especially those involving message-passing, to enable better scalability on multi-node hardware architectures\cite{johansson2022micron, musaelian2023scaling, jia2020pushing}. MLFF models with charge prediction capabilities\cite{deng2023chgnet} can also be interfaced for better adaptivity to the off-stoichiometric IDESs, although the current generation scheme already involves a mechanism to ensure minimum stoichiometric equilibrium requirements, as discussed in \textbf{Section S1} of Supplementary Information. Also, other quantum chemical approaches such as CCSD(T) which is the “gold standard” in accurately describing the potential energy surface but scales poorly in $O(N^7)$\cite{purvis1982full}, can be interfaced for a wide range of systems in computational chemistry, physics, biology, and materials science.

In conclusion, the IDEAL algorithm enables efficient and accurate online simulations of large-scale complex atomic systems, expanding opportunities for studying complex physico-chemical processes and reaction mechanisms at the atomic scale. It shows great promise in enhancing our understanding of various systems and phenomena through its combination of efficiency, accuracy, and scalability.

\section{Methods}

\subsection{Molecular Dynamics}
IDEAL is driven by molecular dynamics or any other sampler such as Monte Carlo. Generally, we initiate the process with an atomic configuration and propagate it using an MLFF model. The next frame is predicted along with a prediction of the uncertainty. If the uncertainty surpasses a certain threshold, additional quantum chemical labeling is involved. Here, we use the on-the-fly learned MLFF as the potential energy surface and conduct MD simulations using the interfaces provided by the Atomic Simulation Environment (ASE)\cite{larsen2017atomic} package. A comprehensive introduction to the MD setups is in \textbf{Section S9} in the Supplementary Information.

\subsection{Uncertainty Quantification}
An uncertainty quantification module is used to assess the reliability of the MLFF prediction results for each step of the MD simulations. The selection of the uncertainty module is orthogonal to the IDEAL algorithm as long as the uncertainty is atom-wise and is differentiable with respect to the atomic coordinates. We design the uncertainty module based on the SOAP descriptor\cite{bartok2013representing} and Mahalanobis distance\cite{mclachlan1999mahalanobis} since the SOAP descriptor effectively converts local environments into vector representations while the Mahalanobis distance serves as a widely used metric for quantifying the distance between a data point and a distribution. We opt for Mahalanobis Distance because it is a reliable, robust, and computationally efficient mathematical tool for implementing uncertainty quantification\cite{filzmoser2008outlier,rousseeuw1990unmasking}. Other methods, such as ensemble models, should also work. We use $D$ to denote the training dataset during the online MD simulation. Each time a high-uncertainty local environment is identified, the corresponding substructure will be generated and will be added to $D$. Let $X=\{x_1,x_2,...,x_M\}$ represent the SOAP features of all local environments that have already been contained in $D$. For a given local environment $\rho_i$, we calculate its SOAP feature $v_i$ using the descriptor. The uncertainty of $\rho_i$ is then determined as: 
$$
\textup{Unc}(\rho_i)=(v_i-\bar x)^T\Sigma^{-1}(v_i-\bar x)
$$
where $\Sigma=\frac{1}{M-1}\sum\limits_{k=1}^{M}(x_k-\bar x)(x_k-\bar x)^T, \bar x=\frac{1}{M}\sum\limits_{k=1}^{M}x_k$.

Additionally, for any structure $\Omega$, the uncertainty of the entire structure, denoted as $\textup{Unc}(\Omega)$, is computed as the aggregation of the $\textup{Unc}(\rho_i)$ values for all $\rho_i$ contained in $\Omega$. To expedite uncertainty computation, we utilize an acceleration technique that mitigates computational complexity. Detailed implementation specifics are provided in \textbf{Section S10} of the Supplementary Information.

\subsection{Quantum Chemical Labeling}
Quantum chemical labeling follows the identification of uncertain local environments and the generation of IDESs. For each IDES, calculations of energy, forces, and, when applicable, stresses are performed using quantum chemical methods. In the current implementation, we use plane-wave DFT with a periodic boundary condition. Except for DFT, some alternative approaches, such as Coupled Cluster with Single, Double, and Triple excitations (CCSD(T))\cite{purvis1982full}, and Quantum Monte Carlo (QMC)\cite{austin2012quantum} can also be employed in this labeling process. Since multiple IDESs are identified for each simulation step, these quantum chemical computations can be further parallelized by distributing them to different computation nodes with high scalability. The experiments presented in \textbf{Section 2} were conducted using the non-parallel version of the IDEAL algorithm. Details of our DFT setups and the parallel version of the IDEAL algorithm, including a comprehensive discussion on its scalability concerning computational resources, can be found in \textbf{Section S11} of the Supplementary Information.

\subsection{The MLFF model and the Training of MLFF}
During the simulation, an MLFF is continuously maintained and updated. The choice of MLFF architecture is, in principle, independent of the IDEAL algorithm. In this study, we opted for M3GNet\cite{chen2022universal}, a graph neural network-based machine learning model known for its ability to provide both rapid and accurate force field predictions. An in-house version of this model was developed using PyTorch. 

During each step of the online MLFF training, we update the model by propagating energy, force, and stress losses computed from a mixture of substructures sampled from the current frame and a pool of previously accumulated samples. This strategy is employed to mitigate catastrophic forgetting that can occur with constant model updates. Specifically, we employ importance sampling to select samples, thereby enhancing training efficiency. For detailed information regarding the training setups, please refer to \textbf{Section S12} in the Supplementary Information.

\section{Notations}
Here we list the notations we use in our paper. 

\begin{table}[H]
\footnotesize
\centering
\begin{tabularx}{\textwidth}{XX}
\toprule
{\multirow{2}{*}{\textbf{Notation}}} & {\multirow{2}{*}{\textbf{Meaning}}} \\
{} & {}  \\
\midrule
$D$ & the dynamically maintained training dataset \\
$\Omega$ & the entire structure \\
$\rho_i$ & the local environment centered around $i$-th atom \\
$\chi_i$ & the corresponding substructure of $\rho_i$ na\"ively cut out from $\Omega$ \\
$\Tilde{\chi_i}$ & the corresponding substructure of $\rho_i$ generated after the first step of in-distribution substructure embedding without uncertainty-driven optimization \\
$\Tilde{\chi_i}^*$ & the corresponding in-distribution embedded substructure of $\rho_i$ generated in IDEAL algorithm\\
$\text{Unc()}$ & the uncertainty module \\
$X=\{x_1, x_2\, ..., x_M\}$ & the local environments already been contained in training dataset $D$ \\
$\bar{x}$ & the mean of $x_i$ in $X$ \\
$\Sigma$ & the covariance matrix of $X$ \\
$v_i$ & the SOAP feature of $\rho_i$ \\
$\text{MAE}_\text{E}$ & the mean absolute error on energies \\
$\text{MAE}_\text{F}$ & the mean absolute error on forces \\
$\text{MAE}_\text{S}$ & the mean absolute error on stresses \\
$t_{\text{entire}}$ & the time cost of a single DFT calculation of the entire structure $\Omega$ \\
$t_{\text{sub}}$ & the time cost of a single DFT calculation of a substructure \\ 
$L$ & the number of high-uncertainty local environments ed in the entire structure with a given uncertainty threshold \\
$N$ & the number of atoms contained in the entire structure \\
$m$ & the number of atoms contained in the substructure \\
\bottomrule
\end{tabularx}
\caption{\textbf{Notations and their corresponding meanings used in our paper.}}
\label{tab:label}
\end{table}

\section{Data Availability}
The DFT-labeled initial structures for the training of MLFF\textsubscript{init}, the structures collected during online simulation using IDEAL, and the test dataset to benchmark MLFF accuracy for Li, Al\textsubscript{2}O\textsubscript{3}, and Haber-Bosch reaction are available upon request.

\section{Acknowledgements}
We thank Tian Xie, Jake Smith, Bichlien Nguyen, Ryota Tomioka, Robert Pinsler, Claudio Zeni, Karin Strauss, and Tie-Yan Liu for insightful discussions; Peggy Dai for managing the project; and colleagues at Microsoft for their encouragement and support.
\section{Author information}

\subsection{Author contributions}
Z.Lu and Y.Zhou led the research. L.Kong, Z.Lu, and Y.Zhou conceived the project. L.Kong and Y.Zhou developed the active learning workflow and model training pipeline. Z.Lu, L.Sun, J.A.G.Torres, and N.Artrith developed data and analytics systems. L.Kong, J.Li, Y.Zhou, and Z.Lu conducted simulations. C.Chen, H.Yang, and H.Hao contributed technical advice and ideas. Z.Lu, L.Kong, and Y.Zhou wrote the paper with the inputs from all authors.

\subsection{Corresponding authors}
Correspondence to \href{zihenglu@microsoft.com}{Ziheng Lu} and \href{yichi.zhou@microsoft.com}{Yichi Zhou}.
\bibliographystyle{unsrt}
\bibliography{ms}

\begin{thebibliography}{10}

\bibitem{car1985unified}
Roberto Car and Mark Parrinello.
\newblock Unified approach for molecular dynamics and density-functional
  theory.
\newblock {\em Physical review letters}, 55(22):2471, 1985.

\bibitem{alfe1999melting}
D~Alfe, MJ~Gillan, and GD~Price.
\newblock The melting curve of iron at the pressures of the earth's core from
  ab initio calculations.
\newblock {\em Nature}, 401(6752):462--464, 1999.

\bibitem{li2022hydrogen}
Peng Li, Yaling Jiang, Youcheng Hu, Yana Men, Yuwen Liu, Wenbin Cai, and
  Shengli Chen.
\newblock Hydrogen bond network connectivity in the electric double layer
  dominates the kinetic ph effect in hydrogen electrocatalysis on pt.
\newblock {\em Nature Catalysis}, 5(10):900--911, 2022.

\bibitem{wang2014discovering}
Lee-Ping Wang, Alexey Titov, Robert McGibbon, Fang Liu, Vijay~S Pande, and
  Todd~J Mart{\'\i}nez.
\newblock Discovering chemistry with an ab initio nanoreactor.
\newblock {\em Nature chemistry}, 6(12):1044--1048, 2014.

\bibitem{hegedus2008microscopic}
J~Heged{\"u}s and SR~Elliott.
\newblock Microscopic origin of the fast crystallization ability of ge--sb--te
  phase-change memory materials.
\newblock {\em Nature materials}, 7(5):399--405, 2008.

\bibitem{ye2021probing}
Zifan Ye, Aleksander Prominski, Bozhi Tian, and Giulia Galli.
\newblock Probing the electronic properties of the electrified silicon/water
  interface by combining simulations and experiments.
\newblock {\em Proceedings of the National Academy of Sciences},
  118(46):e2114929118, 2021.

\bibitem{ye2022photoelectron}
Zifan Ye, Cunzhi Zhang, and Giulia Galli.
\newblock Photoelectron spectra of water and simple aqueous solutions at
  extreme conditions.
\newblock {\em Faraday Discussions}, 236:352--363, 2022.

\bibitem{hohenberg1964inhomogeneous}
Pierre Hohenberg and Walter Kohn.
\newblock Inhomogeneous electron gas.
\newblock {\em Physical review}, 136(3B):B864, 1964.

\bibitem{kohn1965self}
Walter Kohn and Lu~Jeu Sham.
\newblock Self-consistent equations including exchange and correlation effects.
\newblock {\em Physical review}, 140(4A):A1133, 1965.

\bibitem{szabo2012modern}
Attila Szabo and Neil~S Ostlund.
\newblock {\em Modern quantum chemistry: introduction to advanced electronic
  structure theory}.
\newblock Courier Corporation, 2012.

\bibitem{martin2020electronic}
Richard~M Martin.
\newblock {\em Electronic structure: basic theory and practical methods}.
\newblock Cambridge university press, 2020.

\bibitem{behler2007generalized}
J{\"o}rg Behler and Michele Parrinello.
\newblock Generalized neural-network representation of high-dimensional
  potential-energy surfaces.
\newblock {\em Physical review letters}, 98(14):146401, 2007.

\bibitem{shapeev2016moment}
Alexander~V Shapeev.
\newblock Moment tensor potentials: A class of systematically improvable
  interatomic potentials.
\newblock {\em Multiscale Modeling \& Simulation}, 14(3):1153--1173, 2016.

\bibitem{musaelian2023learning}
Albert Musaelian, Simon Batzner, Anders Johansson, Lixin Sun, Cameron~J Owen,
  Mordechai Kornbluth, and Boris Kozinsky.
\newblock Learning local equivariant representations for large-scale atomistic
  dynamics.
\newblock {\em Nature Communications}, 14(1):579, 2023.

\bibitem{chen2019graph}
Chi Chen, Weike Ye, Yunxing Zuo, Chen Zheng, and Shyue~Ping Ong.
\newblock Graph networks as a universal machine learning framework for
  molecules and crystals.
\newblock {\em Chemistry of Materials}, 31(9):3564--3572, 2019.

\bibitem{chen2022universal}
Chi Chen and Shyue~Ping Ong.
\newblock A universal graph deep learning interatomic potential for the
  periodic table.
\newblock {\em Nature Computational Science}, 2(11):718--728, 2022.

\bibitem{park2021accurate}
Cheol~Woo Park, Mordechai Kornbluth, Jonathan Vandermause, Chris Wolverton,
  Boris Kozinsky, and Jonathan~P Mailoa.
\newblock Accurate and scalable graph neural network force field and molecular
  dynamics with direct force architecture.
\newblock {\em npj Computational Materials}, 7(1):73, 2021.

\bibitem{schutt2017schnet}
Kristof Sch{\"u}tt, Pieter-Jan Kindermans, Huziel~Enoc Sauceda~Felix, Stefan
  Chmiela, Alexandre Tkatchenko, and Klaus-Robert M{\"u}ller.
\newblock Schnet: A continuous-filter convolutional neural network for modeling
  quantum interactions.
\newblock {\em Advances in neural information processing systems}, 30, 2017.

\bibitem{wang2023visnet}
Yusong Wang, Shaoning Li, Xinheng He, Mingyu Li, Zun Wang, Nanning Zheng, Bin
  Shao, Tong Wang, and Tie-Yan Liu.
\newblock Visnet: an equivariant geometry-enhanced graph neural network with
  vector-scalar interactive message passing for molecules, 2023.

\bibitem{zhang2018deep}
Linfeng Zhang, Jiequn Han, Han Wang, Roberto Car, and EJPRL Weinan.
\newblock Deep potential molecular dynamics: a scalable model with the accuracy
  of quantum mechanics.
\newblock {\em Physical review letters}, 120(14):143001, 2018.

\bibitem{batzner20223}
Simon Batzner, Albert Musaelian, Lixin Sun, Mario Geiger, Jonathan~P Mailoa,
  Mordechai Kornbluth, Nicola Molinari, Tess~E Smidt, and Boris Kozinsky.
\newblock E (3)-equivariant graph neural networks for data-efficient and
  accurate interatomic potentials.
\newblock {\em Nature communications}, 13(1):2453, 2022.

\bibitem{xie2021bayesian}
Yu~Xie, Jonathan Vandermause, Lixin Sun, Andrea Cepellotti, and Boris Kozinsky.
\newblock Bayesian force fields from active learning for simulation of
  inter-dimensional transformation of stanene.
\newblock {\em npj Computational Materials}, 7(1):40, 2021.

\bibitem{liao2022equiformer}
Yi-Lun Liao and Tess Smidt.
\newblock Equiformer: Equivariant graph attention transformer for 3d atomistic
  graphs.
\newblock {\em arXiv preprint arXiv:2206.11990}, 2022.

\bibitem{johansson2022micron}
Anders Johansson, Yu~Xie, Cameron~J Owen, Jin~Soo Lim, Lixin Sun, Jonathan
  Vandermause, and Boris Kozinsky.
\newblock Micron-scale heterogeneous catalysis with bayesian force fields from
  first principles and active learning.
\newblock {\em arXiv preprint arXiv:2204.12573}, 2022.

\bibitem{musaelian2023scaling}
Albert Musaelian, Anders Johansson, Simon Batzner, and Boris Kozinsky.
\newblock Scaling the leading accuracy of deep equivariant models to
  biomolecular simulations of realistic size.
\newblock {\em arXiv preprint arXiv:2304.10061}, 2023.

\bibitem{jia2020pushing}
Weile Jia, Han Wang, Mohan Chen, Denghui Lu, Lin Lin, Roberto Car, E~Weinan,
  and Linfeng Zhang.
\newblock Pushing the limit of molecular dynamics with ab initio accuracy to
  100 million atoms with machine learning.
\newblock In {\em SC20: International conference for high performance
  computing, networking, storage and analysis}, pages 1--14. IEEE, 2020.

\bibitem{unke2021machine}
Oliver~T Unke, Stefan Chmiela, Huziel~E Sauceda, Michael Gastegger, Igor
  Poltavsky, Kristof~T Sch{\"u}tt, Alexandre Tkatchenko, and Klaus-Robert
  M{\"u}ller.
\newblock Machine learning force fields.
\newblock {\em Chemical Reviews}, 121(16):10142--10186, 2021.

\bibitem{bartok2010gaussian}
Albert~P Bart{\'o}k, Mike~C Payne, Risi Kondor, and G{\'a}bor Cs{\'a}nyi.
\newblock Gaussian approximation potentials: The accuracy of quantum mechanics,
  without the electrons.
\newblock {\em Physical review letters}, 104(13):136403, 2010.

\bibitem{thompson2015spectral}
Aidan~P Thompson, Laura~P Swiler, Christian~R Trott, Stephen~M Foiles, and
  Garritt~J Tucker.
\newblock Spectral neighbor analysis method for automated generation of
  quantum-accurate interatomic potentials.
\newblock {\em Journal of Computational Physics}, 285:316--330, 2015.

\bibitem{drautz2019atomic}
Ralf Drautz.
\newblock Atomic cluster expansion for accurate and transferable interatomic
  potentials.
\newblock {\em Physical Review B}, 99(1):014104, 2019.

\bibitem{fu2022forces}
Xiang Fu, Zhenghao Wu, Wujie Wang, Tian Xie, Sinan Keten, Rafael
  Gomez-Bombarelli, and Tommi Jaakkola.
\newblock Forces are not enough: Benchmark and critical evaluation for machine
  learning force fields with molecular simulations.
\newblock {\em arXiv preprint arXiv:2210.07237}, 2022.

\bibitem{vandermause2020fly}
Jonathan Vandermause, Steven~B Torrisi, Simon Batzner, Yu~Xie, Lixin Sun,
  Alexie~M Kolpak, and Boris Kozinsky.
\newblock On-the-fly active learning of interpretable bayesian force fields for
  atomistic rare events.
\newblock {\em npj Computational Materials}, 6(1):20, 2020.

\bibitem{vandermause2022active}
Jonathan Vandermause, Yu~Xie, Jin~Soo Lim, Cameron~J Owen, and Boris Kozinsky.
\newblock Active learning of reactive bayesian force fields applied to
  heterogeneous catalysis dynamics of h/pt.
\newblock {\em Nature Communications}, 13(1):5183, 2022.

\bibitem{gubaev2019accelerating}
Konstantin Gubaev, Evgeny~V Podryabinkin, Gus~LW Hart, and Alexander~V Shapeev.
\newblock Accelerating high-throughput searches for new alloys with active
  learning of interatomic potentials.
\newblock {\em Computational Materials Science}, 156:148--156, 2019.

\bibitem{jinnouchi2020fly}
Ryosuke Jinnouchi, Kazutoshi Miwa, Ferenc Karsai, Georg Kresse, and Ryoji
  Asahi.
\newblock On-the-fly active learning of interatomic potentials for large-scale
  atomistic simulations.
\newblock {\em The Journal of Physical Chemistry Letters}, 11(17):6946--6955,
  2020.

\bibitem{kulichenko2023uncertainty}
Maksim Kulichenko, Kipton Barros, Nicholas Lubbers, Ying~Wai Li, Richard
  Messerly, Sergei Tretiak, Justin~S Smith, and Benjamin Nebgen.
\newblock Uncertainty-driven dynamics for active learning of interatomic
  potentials.
\newblock {\em Nature Computational Science}, 3(3):230--239, 2023.

\bibitem{podryabinkin2017active}
Evgeny~V Podryabinkin and Alexander~V Shapeev.
\newblock Active learning of linearly parametrized interatomic potentials.
\newblock {\em Computational Materials Science}, 140:171--180, 2017.

\bibitem{podryabinkin2019accelerating}
Evgeny~V Podryabinkin, Evgeny~V Tikhonov, Alexander~V Shapeev, and Artem~R
  Oganov.
\newblock Accelerating crystal structure prediction by machine-learning
  interatomic potentials with active learning.
\newblock {\em Physical Review B}, 99(6):064114, 2019.

\bibitem{schran2020committee}
Christoph Schran, Krystof Brezina, and Ondrej Marsalek.
\newblock Committee neural network potentials control generalization errors and
  enable active learning.
\newblock {\em The Journal of Chemical Physics}, 153(10), 2020.

\bibitem{smith2018less}
Justin~S Smith, Ben Nebgen, Nicholas Lubbers, Olexandr Isayev, and Adrian~E
  Roitberg.
\newblock Less is more: Sampling chemical space with active learning.
\newblock {\em The Journal of chemical physics}, 148(24), 2018.

\bibitem{wilson2022batch}
Nathan Wilson, Daniel Willhelm, Xiaoning Qian, Raymundo Arr{\'o}yave, and
  Xiaofeng Qian.
\newblock Batch active learning for accelerating the development of interatomic
  potentials.
\newblock {\em Computational Materials Science}, 208:111330, 2022.

\bibitem{xie2023uncertainty}
Yu~Xie, Jonathan Vandermause, Senja Ramakers, Nakib~H Protik, Anders Johansson,
  and Boris Kozinsky.
\newblock Uncertainty-aware molecular dynamics from bayesian active learning
  for phase transformations and thermal transport in sic.
\newblock {\em npj Computational Materials}, 9(1):36, 2023.

\bibitem{zhang2019active}
Linfeng Zhang, De-Ye Lin, Han Wang, Roberto Car, and E~Weinan.
\newblock Active learning of uniformly accurate interatomic potentials for
  materials simulation.
\newblock {\em Physical Review Materials}, 3(2):023804, 2019.

\bibitem{kuhne2020cp2k}
Thomas~D K{\"u}hne, Marcella Iannuzzi, Mauro Del~Ben, Vladimir~V Rybkin,
  Patrick Seewald, Frederick Stein, Teodoro Laino, Rustam~Z Khaliullin, Ole
  Sch{\"u}tt, Florian Schiffmann, et~al.
\newblock Cp2k: An electronic structure and molecular dynamics software
  package-quickstep: Efficient and accurate electronic structure calculations.
\newblock {\em The Journal of Chemical Physics}, 152(19), 2020.

\bibitem{giannozzi2020quantum}
Paolo Giannozzi, Oscar Baseggio, Pietro Bonf{\`a}, Davide Brunato, Roberto Car,
  Ivan Carnimeo, Carlo Cavazzoni, Stefano De~Gironcoli, Pietro Delugas,
  Fabrizio Ferrari~Ruffino, et~al.
\newblock Quantum espresso toward the exascale.
\newblock {\em The Journal of chemical physics}, 152(15), 2020.

\bibitem{smith2020psi4}
Daniel~GA Smith, Lori~A Burns, Andrew~C Simmonett, Robert~M Parrish, Matthew~C
  Schieber, Raimondas Galvelis, Peter Kraus, Holger Kruse, Roberto Di~Remigio,
  Asem Alenaizan, et~al.
\newblock Psi4 1.4: Open-source software for high-throughput quantum chemistry.
\newblock {\em The Journal of chemical physics}, 152(18), 2020.

\bibitem{podryabinkin2023mlip}
Evgeny Podryabinkin, Kamil Garifullin, Alexander Shapeev, and Ivan Novikov.
\newblock Mlip-3: Active learning on atomic environments with moment tensor
  potentials.
\newblock {\em arXiv preprint arXiv:2304.13144}, 2023.

\bibitem{pickard2011ab}
Chris~J Pickard and RJ~Needs.
\newblock Ab initio random structure searching.
\newblock {\em Journal of Physics: Condensed Matter}, 23(5):053201, 2011.

\bibitem{cheng2017toward}
Xin-Bing Cheng, Rui Zhang, Chen-Zi Zhao, and Qiang Zhang.
\newblock Toward safe lithium metal anode in rechargeable batteries: a review.
\newblock {\em Chemical reviews}, 117(15):10403--10473, 2017.

\bibitem{sutar2010ring}
Alekha~Kumar Sutar, Tungabidya Maharana, Saikat Dutta, Chi-Tien Chen, and
  Chu-Chieh Lin.
\newblock Ring-opening polymerization by lithium catalysts: an overview.
\newblock {\em Chemical Society Reviews}, 39(5):1724--1746, 2010.

\bibitem{han2017negating}
Xiaogang Han, Yunhui Gong, Kun Fu, Xingfeng He, Gregory~T Hitz, Jiaqi Dai, Alex
  Pearse, Boyang Liu, Howard Wang, Gary Rubloff, et~al.
\newblock Negating interfacial impedance in garnet-based solid-state li metal
  batteries.
\newblock {\em Nature materials}, 16(5):572--579, 2017.

\bibitem{rodrigues2017materials}
Marco-Tulio~F Rodrigues, Ganguli Babu, Hemtej Gullapalli, Kaushik Kalaga,
  Farheen~N Sayed, Keiko Kato, Jarin Joyner, and Pulickel~M Ajayan.
\newblock A materials perspective on li-ion batteries at extreme temperatures.
\newblock {\em nature energy}, 2(8):1--14, 2017.

\bibitem{hinton2002stochastic}
Geoffrey~E Hinton and Sam Roweis.
\newblock Stochastic neighbor embedding.
\newblock {\em Advances in neural information processing systems}, 15, 2002.

\bibitem{bartok2013representing}
Albert~P Bart{\'o}k, Risi Kondor, and G{\'a}bor Cs{\'a}nyi.
\newblock On representing chemical environments.
\newblock {\em Physical Review B}, 87(18):184115, 2013.

\bibitem{perdew1996generalized}
John~P Perdew, Kieron Burke, and Matthias Ernzerhof.
\newblock Generalized gradient approximation made simple.
\newblock {\em Physical review letters}, 77(18):3865, 1996.

\bibitem{shi2010decreasing}
Xiaomeng Shi and Jian Luo.
\newblock Decreasing the grain boundary diffusivity in binary alloys with
  increasing temperature.
\newblock {\em Physical review letters}, 105(23):236102, 2010.

\bibitem{straumal2014grain}
BB~Straumal, A~Korneva, O~Kogtenkova, L~Kurmanaeva, P~Zi{\k{e}}ba,
  A~Wierzbicka-Miernik, SN~Zhevnenko, and B~Baretzky.
\newblock Grain boundary wetting and premelting in the cu--co alloys.
\newblock {\em Journal of alloys and compounds}, 615:S183--S187, 2014.

\bibitem{frolov2013structural}
Timofey Frolov, David~L Olmsted, Mark Asta, and Yuri Mishin.
\newblock Structural phase transformations in metallic grain boundaries.
\newblock {\em Nature communications}, 4(1):1899, 2013.

\bibitem{deng2022high}
Bing Deng, Paul~A Advincula, Duy~Xuan Luong, Jingan Zhou, Boyu Zhang, Zhe Wang,
  Emily~A McHugh, Jinhang Chen, Robert~A Carter, Carter Kittrell, et~al.
\newblock High-surface-area corundum nanoparticles by resistive hotspot-induced
  phase transformation.
\newblock {\em Nature Communications}, 13(1):5027, 2022.

\bibitem{macfarlane2020roadmap}
Douglas~R MacFarlane, Pavel~V Cherepanov, Jaecheol Choi, Bryan~HR Suryanto,
  Rebecca~Y Hodgetts, Jacinta~M Bakker, Federico M~Ferrero Vallana, and
  Alexandr~N Simonov.
\newblock A roadmap to the ammonia economy.
\newblock {\em Joule}, 4(6):1186--1205, 2020.

\bibitem{goedecker1999linear}
Stefan Goedecker.
\newblock Linear scaling electronic structure methods.
\newblock {\em Reviews of Modern Physics}, 71(4):1085, 1999.

\bibitem{deng2023chgnet}
Bowen Deng, Peichen Zhong, KyuJung Jun, Janosh Riebesell, Kevin Han,
  Christopher~J Bartel, and Gerbrand Ceder.
\newblock Chgnet as a pretrained universal neural network potential for
  charge-informed atomistic modelling.
\newblock {\em Nature Machine Intelligence}, pages 1--11, 2023.

\bibitem{purvis1982full}
George~D Purvis~III and Rodney~J Bartlett.
\newblock A full coupled-cluster singles and doubles model: The inclusion of
  disconnected triples.
\newblock {\em The Journal of Chemical Physics}, 76(4):1910--1918, 1982.

\bibitem{larsen2017atomic}
Ask~Hjorth Larsen, Jens~J{\o}rgen Mortensen, Jakob Blomqvist, Ivano~E Castelli,
  Rune Christensen, Marcin Du{\l}ak, Jesper Friis, Michael~N Groves, Bj{\o}rk
  Hammer, Cory Hargus, et~al.
\newblock The atomic simulation environment—a python library for working with
  atoms.
\newblock {\em Journal of Physics: Condensed Matter}, 29(27):273002, 2017.

\bibitem{mclachlan1999mahalanobis}
Goeffrey~J McLachlan.
\newblock Mahalanobis distance.
\newblock {\em Resonance}, 4(6):20--26, 1999.

\bibitem{filzmoser2008outlier}
Peter Filzmoser, Ricardo Maronna, and Mark Werner.
\newblock Outlier identification in high dimensions.
\newblock {\em Computational statistics \& data analysis}, 52(3):1694--1711,
  2008.

\bibitem{rousseeuw1990unmasking}
Peter~J Rousseeuw and Bert~C Van~Zomeren.
\newblock Unmasking multivariate outliers and leverage points.
\newblock {\em Journal of the American Statistical association},
  85(411):633--639, 1990.

\bibitem{austin2012quantum}
Brian~M Austin, Dmitry~Yu Zubarev, and William~A Lester~Jr.
\newblock Quantum monte carlo and related approaches.
\newblock {\em Chemical reviews}, 112(1):263--288, 2012.

\end{thebibliography}


\begin{thebibliography}{10}

\bibitem{hirel2015atomsk}
Pierre Hirel.
\newblock Atomsk: A tool for manipulating and converting atomic data files.
\newblock {\em Computer Physics Communications}, 197:212--219, 2015.

\bibitem{larsen2017atomic}
Ask~Hjorth Larsen, Jens~J{\o}rgen Mortensen, Jakob Blomqvist, Ivano~E Castelli,
  Rune Christensen, Marcin Du{\l}ak, Jesper Friis, Michael~N Groves, Bj{\o}rk
  Hammer, Cory Hargus, et~al.
\newblock The atomic simulation environment—a python library for working with
  atoms.
\newblock {\em Journal of Physics: Condensed Matter}, 29(27):273002, 2017.

\bibitem{frenkel2023understanding}
Daan Frenkel and Berend Smit.
\newblock {\em Understanding molecular simulation: from algorithms to
  applications}.
\newblock Elsevier, 2023.

\bibitem{allen2017computer}
Michael~P Allen and Dominic~J Tildesley.
\newblock {\em Computer simulation of liquids}.
\newblock Oxford university press, 2017.

\bibitem{berendsen1984molecular}
Herman~JC Berendsen, JPM~van Postma, Wilfred~F Van~Gunsteren, ARHJ DiNola, and
  Jan~R Haak.
\newblock Molecular dynamics with coupling to an external bath.
\newblock {\em The Journal of chemical physics}, 81(8):3684--3690, 1984.

\bibitem{hunenberger1999ewald}
Philippe~H H{\"u}nenberger and J~Andrew McCammon.
\newblock Ewald artifacts in computer simulations of ionic solvation and
  ion--ion interaction: a continuum electrostatics study.
\newblock {\em The Journal of chemical physics}, 110(4):1856--1872, 1999.

\bibitem{bartok2013representing}
Albert~P Bart{\'o}k, Risi Kondor, and G{\'a}bor Cs{\'a}nyi.
\newblock On representing chemical environments.
\newblock {\em Physical Review B}, 87(18):184115, 2013.

\bibitem{mclachlan1999mahalanobis}
Goeffrey~J McLachlan.
\newblock Mahalanobis distance.
\newblock {\em Resonance}, 4(6):20--26, 1999.

\bibitem{himanen2020dscribe}
Lauri Himanen, Marc~OJ J{\"a}ger, Eiaki~V Morooka, Filippo~Federici Canova,
  Yashasvi~S Ranawat, David~Z Gao, Patrick Rinke, and Adam~S Foster.
\newblock Dscribe: Library of descriptors for machine learning in materials
  science.
\newblock {\em Computer Physics Communications}, 247:106949, 2020.

\bibitem{laakso2023updates}
Jarno Laakso, Lauri Himanen, Henrietta Homm, Eiaki~V Morooka, Marc~OJ
  J{\"a}ger, Milica Todorovi{\'c}, and Patrick Rinke.
\newblock Updates to the dscribe library: New descriptors and derivatives.
\newblock {\em The Journal of Chemical Physics}, 158(23), 2023.

\bibitem{kresse1996efficiency}
Georg Kresse and J{\"u}rgen Furthm{\"u}ller.
\newblock Efficiency of ab-initio total energy calculations for metals and
  semiconductors using a plane-wave basis set.
\newblock {\em Computational materials science}, 6(1):15--50, 1996.

\bibitem{chen2022universal}
Chi Chen and Shyue~Ping Ong.
\newblock A universal graph deep learning interatomic potential for the
  periodic table.
\newblock {\em Nature Computational Science}, 2(11):718--728, 2022.

\bibitem{loshchilov2015online}
Ilya Loshchilov and Frank Hutter.
\newblock Online batch selection for faster training of neural networks.
\newblock {\em arXiv preprint arXiv:1511.06343}, 2015.

\bibitem{katharopoulos2018not}
Angelos Katharopoulos and Fran{\c{c}}ois Fleuret.
\newblock Not all samples are created equal: Deep learning with importance
  sampling.
\newblock In {\em International conference on machine learning}, pages
  2525--2534. PMLR, 2018.

\end{thebibliography}

\end{document}


\maketitle

\section{Determination of Cropping Cell}
During the process of substructure cropping and embedding, nonphysical neighbor relation and the off-stoichiometric composition may be encountered. For example, for the case of Al\textsubscript{2}O\textsubscript{3}, if the cropping boundary is not properly selected, O-O and Al-Al neighbors will be generated instead of the Al-O neighbors in the training set. Such is also reflected by the non-smooth and multi-minima behavior of the uncertainty with respect to the cropping cell boundary. In other words, the introduction of a new atom can result in discontinuities in the uncertainty values. This poses challenges when attempting to directly optimize the uncertainty. To address these issues, we employ a randomized approach where we sample between 20 to 100 cropping cell boundaries and subsequently select the one that yields the lowest uncertainty for further uncertainty-driven optimization. 

\begin{figure}[H] 
  \centering 
  \includegraphics[width=0.8\textwidth]{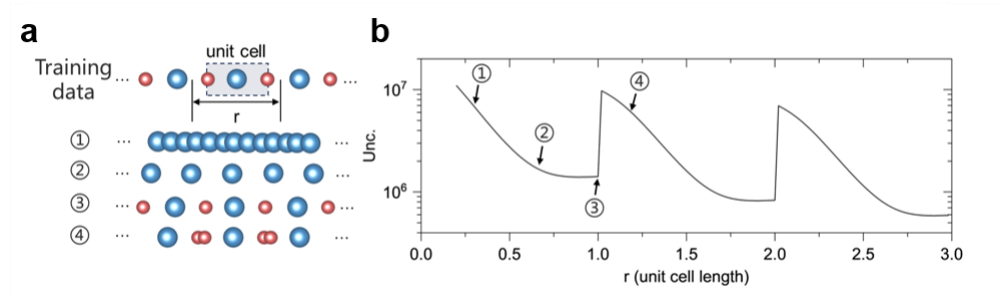}
  \caption{\textbf{Substructures generated under various unit cell sizes. } \textbf{a} Substructures are cropped off from a one-dimensional chain-like structure and different atomic nearest-neighbor relationships are formed by various substructure unit cell boundary sizes. \textbf{b} The uncertainty plot of substructures cropped off from the one-dimensional chain-like structure with varying unit cell lengths.  } 
  \label{fig:example}
\end{figure}

To illustrate the impact of different cropping boundaries on substructure uncertainty, we employed a simplified model featuring a one-dimensional alternating periodic structure composed of aluminum (Al) and oxygen (O). As depicted in \textbf{Supplementary Figure 1}, different cropping boundaries resulted in varying degrees of uncertainty. Specifically, when the cropping boundary intersected specific atoms, it induced alterations in the adjacent atomic environments. For instance, if we considered the Al atom at the center, reducing the cropping length to less than one unit cell resulted in the formation of -Al-Al-Al- chains, deviating from the original alternating Al chains. Therefore, by carefully selecting the cropping boundary, one can generate a structure with controlled uncertainty while preserving the neighboring relations and a rough conservation of the stoichiometry.

\section{Ground state structures of Li and Al\textsubscript{2}O\textsubscript{3}}
The ground state structures of crystalline lithium and Al\textsubscript{2}O\textsubscript{3} were computed by minimizing the energies using Density functional Theory (DFT) and actively learned machine learning forcefields (MLFF\textsubscript{IDEAL}). Specifically, the lattice parameters as well as the atomic positions were left to relax until the forces are smaller than 0.01eV/Å. The results of the relaxed structures are shown in \textbf{Supplementary Table 1}.

\begin{table}[H]
\footnotesize
\centering
\begin{tabularx}{\textwidth}{XXXXXXXX}
\toprule
{\multirow{2}{*}{}} & {\multirow{2}{*}{}} & {\multirow{2}{*}{\textbf{a (Å)}}} & {\multirow{2}{*}{\textbf{b (Å)}}} & {\multirow{2}{*}{\textbf{c (Å)}}} & {\multirow{2}{*}{\textbf{$\alpha$ (º)}}} & {\multirow{2}{*}{\textbf{$\beta$ (º)}}} & {\multirow{2}{*}{\textbf{$\gamma$ (º)}}} \\
{} & {} & {} & {} & {} & {} & {} & {}\\
\midrule
Li\textsubscript{BCC} & DFT & 3.44 & 3.44 & 3.44 & 90 & 90 & 90 \\
 & MLFF\textsubscript{IDEAL} & 3.43 & 3.43 & 3.43 & 90 & 90 & 90 \\
 & Error (\%) & 0.27 & 0.27 & 0.27 & 0 & 0 & 0 \\
\hline
Li\textsubscript{FCC} & DFT & 4.33 & 4.33 & 4.33 & 90 & 90 & 90 \\
 & MLFF\textsubscript{IDEAL} & 4.32 & 4.32 & 4.32 & 90 & 90 & 90 \\
 & Error (\%) & 0.11 & 0.11 & 0.11 & 0 & 0 & 0 \\
\hline
Al\textsubscript{2}O\textsubscript{3} & DFT & 4.81 & 4.81 & 13.12 & 90 & 90 & 120 \\
 & MLFF\textsubscript{IDEAL} & 4.81 & 4.81 & 13.11 & 90 & 90 & 120 \\
 & Error (\%) & 0.001 & 0.001 & 0.07 & 0 & 0 & 0 \\
\bottomrule
\end{tabularx}
\caption{\textbf{Comparison of lattice parameters for ground state structures relaxed with DFT and actively learned MLFFs. }}
\label{tab:label}
\end{table}

\section{Comparison between radial distributions of molten Li}


For lithium metal, comparisons between \textit{ab initio} molecular dynamics (AIMD) and actively learned MLFF (MLFF\textsubscript{IDEAL}) were carried out by a Constant Number, Pressure, and Temperature (NPT) simulation of a BCC Li system at 800K for 50 ps under ambient pressure and the corresponding radial distribution function (RDF) plots of the molten lithium are shown in \textbf{Supplementary Figure 2}.

\begin{figure}[H] 
  \centering 
  \includegraphics[width=0.8\textwidth]{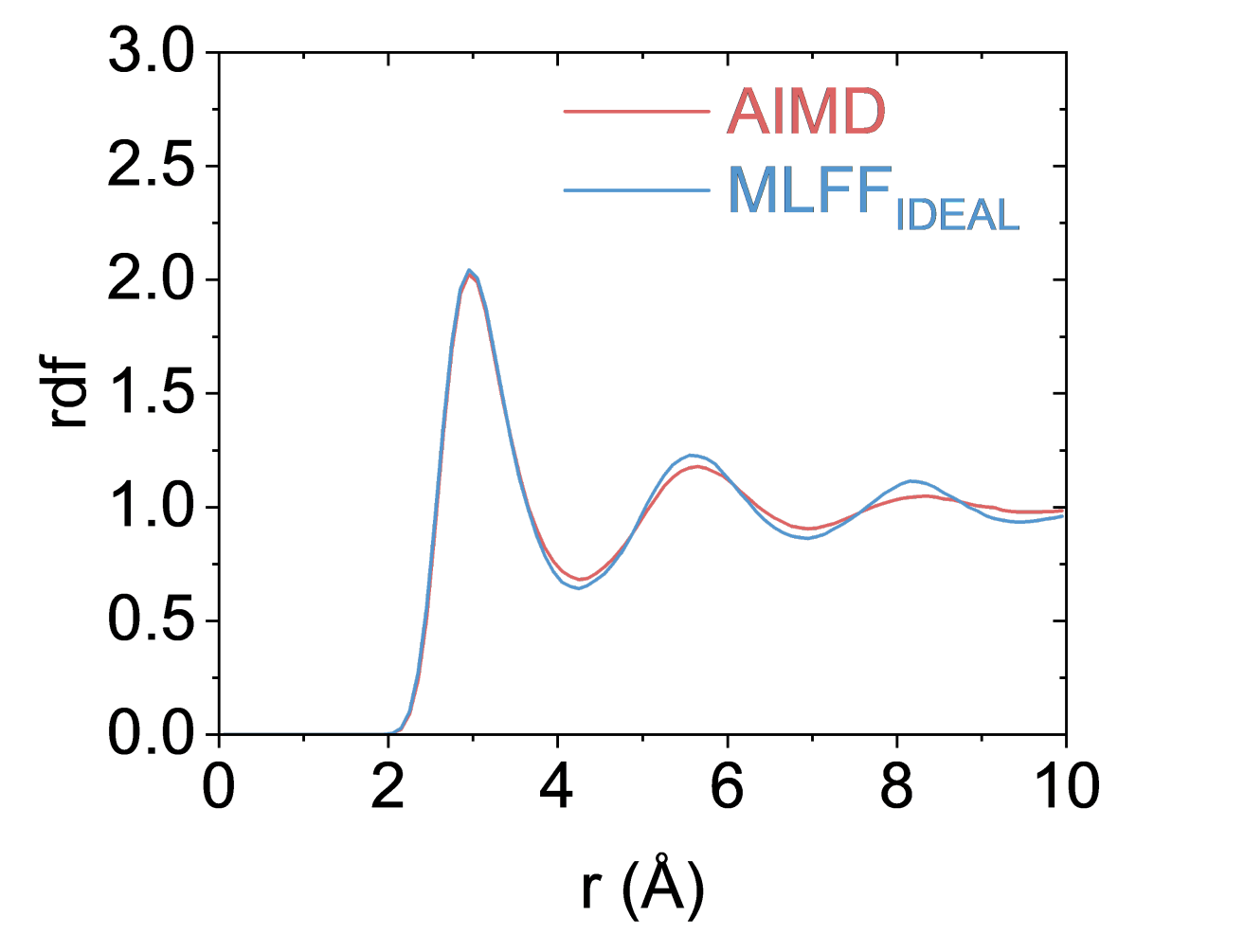}
  \caption{\textbf{Comparison of RDF plots for molten Li simulated using AIMD method and actively learned MLFF. }} 
  \label{fig:example}
\end{figure}

\section{Simulation details of one-million-atom polycrystalline Li }
The IDEAL simulation of the melting of polycrystalline Li was carried out on a structure with 1.02 million atoms. The structure was constructed using Atomsk\cite{hirel2015atomsk} based on a body-centered cubic (BCC) template. A cubic simulation cell with a side length of 28.66 nm is used and 20 randomly-placed crystal particles were included. The system was pre-relaxed using a classical forcefield before carrying out the IDEAL-based MD simulation. We then used the IDEAL algorithm to conduct an online MD simulation on this structure under 600K and saved the structure every 20 fs. It is noteworthy that, to ensure the simulation can be executed on a single 80G-A100 GPU, we reduced the number of model parameters of the MLFF. Specific parameters of the model are listed in \textbf{Section S12}. To validate the accuracy of the model not being affected by shrinking its size, we trained this reduced-parameter MLFF on all the substructure data collected in previous experiments and tested it on 600 Li\textsubscript{512} structures. The MAE\textsubscript{E} and MAE\textsubscript{F} based on the full-size model and reduced-parameter models were 0.004eV/atom, 0.035eV/Å, and 0.022eV/atom, 0.083eV/Å respectively. The test results indicate that, despite a significant reduction in model parameters, the model's accuracy can still be maintained.

The snapshots of the simulation are shown in \textbf{Figure 2i-k} in the main context. A clear grain-boundary-initiated melting is observed. In fact, the grain boundary serves as seeds to the melting process, and a reaction front can be observed propagating from the grain boundary to the crystalline region, finally leading to the full melting of the material. As shown in \textbf{Supplementary Figure 3}, both the RDF and the identification of local symmetry assist the full melting of the sample.  

\begin{figure}[H] 
  \centering 
  \includegraphics[width=\textwidth]{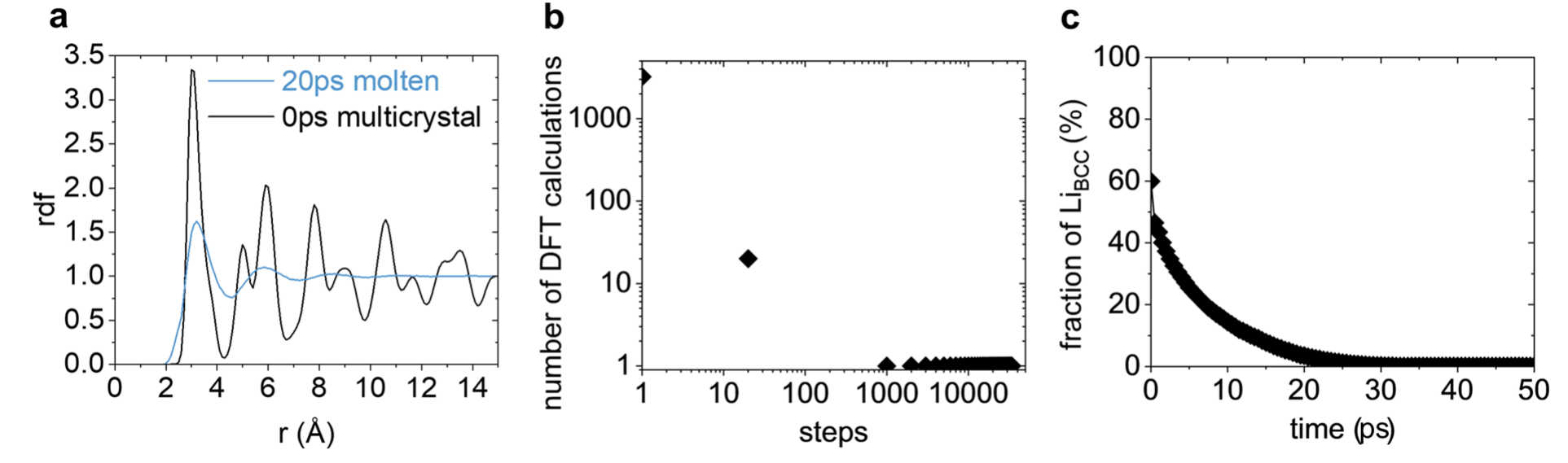}
  \caption{\textbf{Experiment results on one-million-atom polycrystalline Li. } \textbf{a} The RDF plots for multicrystal and molten structure. \textbf{b} The number of DFT calculations invoked during the online simulation. \textbf{c} The fraction of body-centered cubic part in the structure. } 
  \label{fig:example}
\end{figure}

\section{Simulation of melting temperature Al\textsubscript{2}O\textsubscript{3}}

\begin{figure}[H] 
  \centering 
  \includegraphics[width=0.8\textwidth]{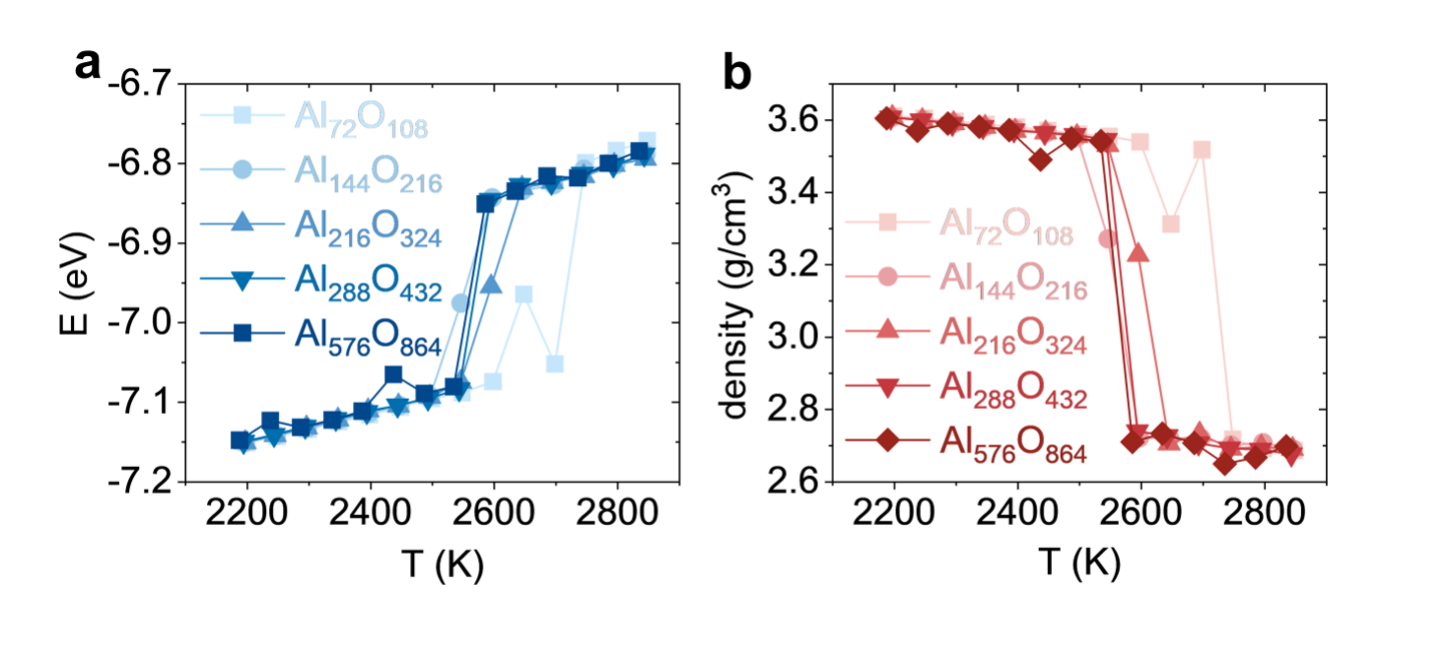}
  \caption{\textbf{Al\textsubscript{2}O\textsubscript{3} melting point determined using MLFF learned with IDEAL algorithm. } \textbf{a} Average potential energy of Al\textsubscript{2}O\textsubscript{3} during an equilibrium MD simulation at temperatures ranging from 2200K to 2850K \textbf{b} Average density of Al\textsubscript{2}O\textsubscript{3} during an equilibrium MD simulation at temperatures ranging from 2200K to 2850K. } 
  \label{fig:example}
\end{figure}

The simulation of Al\textsubscript{2}O\textsubscript{3} bulk phase was carried out based on the R$\bar{3}$m structure under ambient pressure, with the temperature ranging from 2200K to 2850K. As shown in \textbf{Supplementary Figure 4}, we observed discontinuous jumps in both density and per-atom energy at around 2550K, which is the melting point estimated by the machine learning forcefield (MLFF). 

\section{Melting Experiment of Al\textsubscript{2}O\textsubscript{3} Nano-Particle}
The melting simulation of the Al\textsubscript{328}O\textsubscript{492} (Al\textsubscript{2}O\textsubscript{3}) nano-particle assesses the effectiveness of the IDEAL algorithm in MD simulations involving interface structures. The MD simulation started from an Al\textsubscript{2}O\textsubscript{3} nanoparticle containing 820 atoms and used the same initial dataset as the Al\textsubscript{324}O\textsubscript{486} experiment. We heated the structure to 3000 K and simulated the molecular dynamics process for 30 ps. Some snapshots about this melting trajectory were shown in \textbf{Supplementary Figure 5}. 

\begin{figure}[H] 
  \centering
  \includegraphics[width=0.8\textwidth]{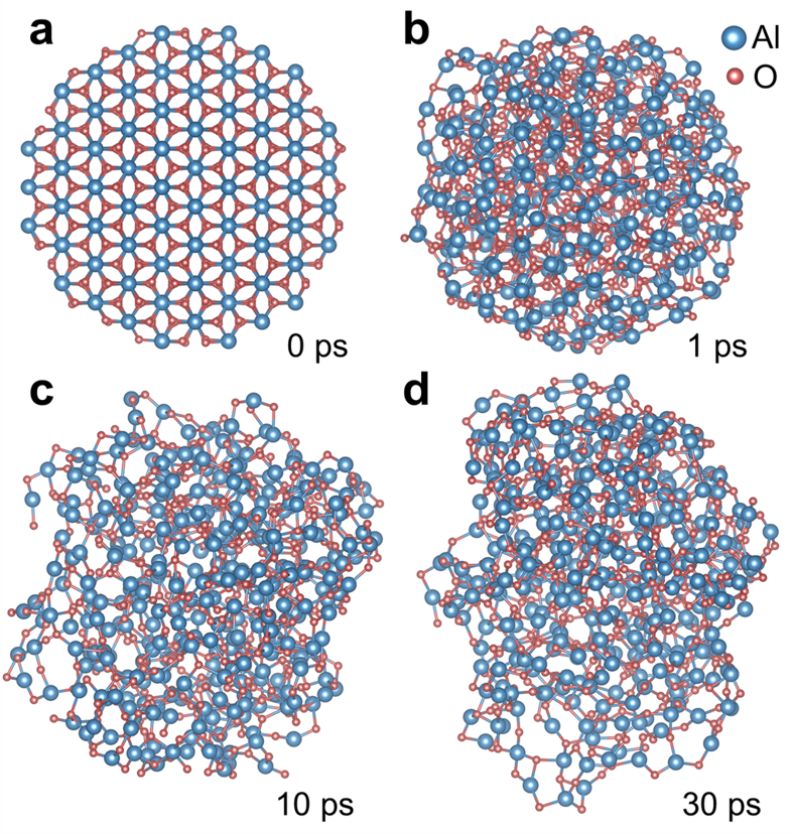}
  \caption{\textbf{Snapshot of Al\textsubscript{2}O\textsubscript{3} nanoparticle melting trajectory at corresponding time. }} 
  \label{fig:example}
\end{figure}

To assess the accuracy of the model trained using the IDEAL algorithm, we generated an MD trajectory of a smaller nanoparticle comprising 481 atoms and sampled 500 structures from this trajectory to create a test set. (We opted not to construct the test set from the MD trajectory of Al\textsubscript{328}O\textsubscript{492}, as performing DFT calculations on such an extensive nanoparticle would be prohibitively expensive. Notably, the primary distinction between nanoparticles of varying sizes lies in their radii, with no significant structural differences; hence, we opted for slightly smaller nano-spheres to create the test set.) Subsequently, we evaluated the model's accuracy on these nanoparticle structures and compared the results with those obtained from a model trained using the initial dataset. The findings are presented in \textbf{Supplementary Table 2}. 

\begin{table}[H]
\footnotesize
\centering
\begin{tabularx}{\textwidth}{XXX}
\toprule
{\multirow{2}{*}{\textbf{Model}}} & {\multirow{2}{*}{\textbf{MAE\textsubscript{E} (eV/atom)}}} & {\multirow{2}{*}{\textbf{MAE\textsubscript{F} (eV/Å)}}} \\
{} &  {} & {} \\
\midrule
MLFF\textsubscript{init} & 0.303 & 0.595 \\
MLFF\textsubscript{IDEAL} & 0.003 & 0.246 \\
\bottomrule
\end{tabularx}
\caption{\textbf{Performance on Al\textsubscript{2}O\textsubscript{3} nanoparticle of MLFF on-the-fly trained by IDEAL compared with initial MLFF. }}
\label{tab:label}
\end{table}

The experimental results demonstrate that the IDEAL algorithm can accomplish rapid and precise MD simulations on large-scale interface structures. This indicates that the algorithm holds the potential to be employed in studying various interface phenomena in material science and related fields.

\section{Simulation details of Haber-Bosch reaction}
Before commencing the Haber-Bosch simulation, we created an initialization dataset by randomly placing N\textsubscript{2} and H\textsubscript{2} molecules, N atoms, and H atoms on a 40-atom Iridium plane and relaxed the system. This dataset with 9000 structures inside helped us initialize the IDEAL algorithm. 

We found that simulating the Haber-Bosch reaction directly from scratch using IDEAL is quite challenging due to the prolonged periods and stringent conditions often required for the dissociation of N\textsubscript{2}. Therefore, as a workaround, we initially placed some pre-dissociated N atoms on the surface of an Iridium nanoparticle structure. Starting from this modified structure, we employed the IDEAL algorithm to simulate the reaction at 1200K. During this simulation, we observed the adsorption, and dissociation of H\textsubscript{2}, as well as the generation and desorption processes of NH, NH\textsubscript{2}, and NH\textsubscript{3}. Regarding the dissociation of N\textsubscript{2}, we constructed an additional initial structure by placing the Iridium nanoparticle in a pure N\textsubscript{2} gas atmosphere and heating it to 7000K. It is noteworthy that, for the N\textsubscript{2} dissociation simulation, we fixed the core of the Iridium nanoparticle in the simulation, allowing only the outermost two layers of atoms on the Iridium nanoparticle to move during the simulation. In this simulation, we were able to observe the adsorption and dissociation of N\textsubscript{2}. All the key steps of Haber-Bosch reaction simulated by IDEAL in these experiments are shown in \textbf{Supplementary Figure 6}.

\begin{figure}[H] 
  \centering 
  \includegraphics[width=\textwidth]{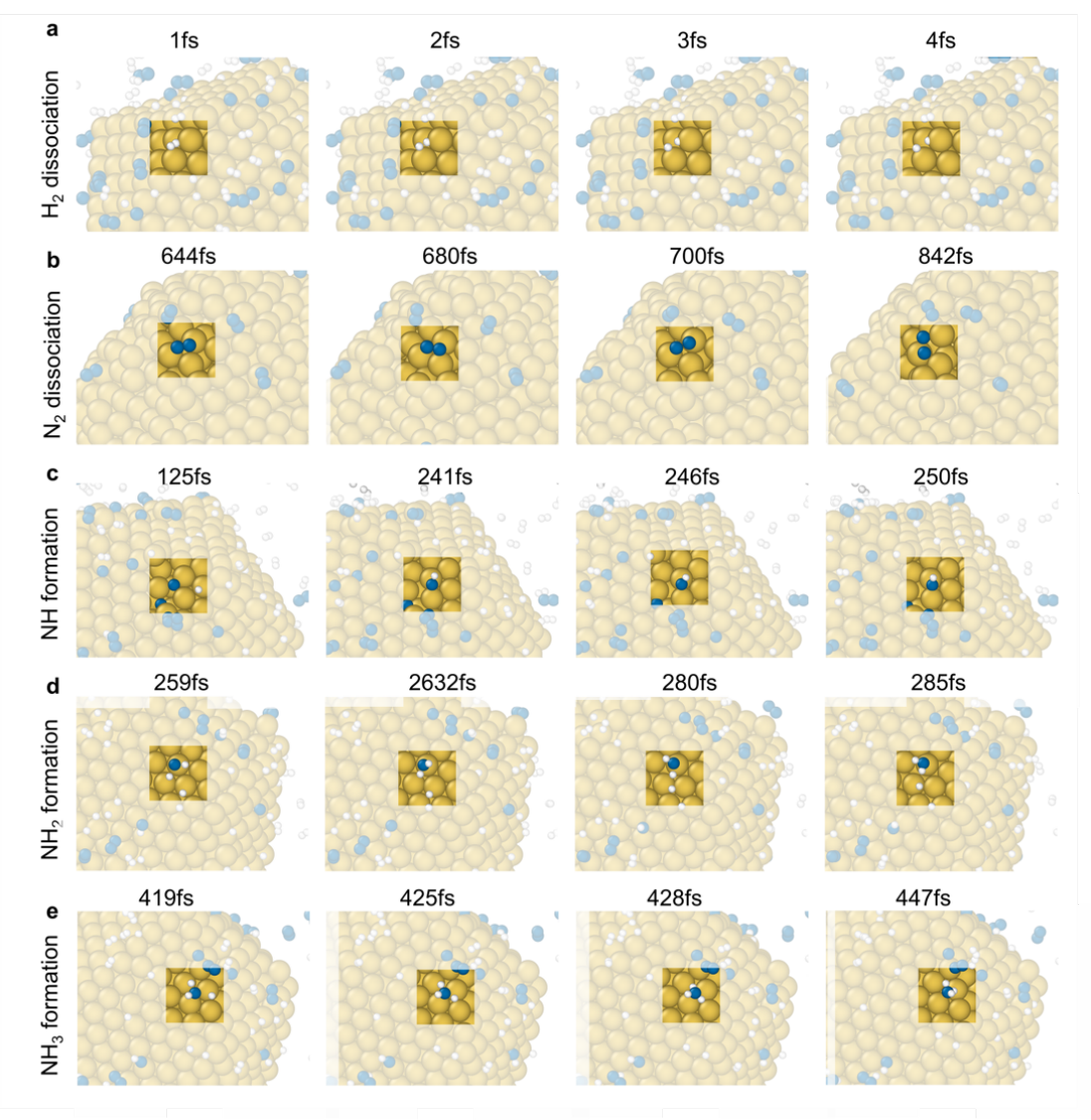}
  \caption{\footnotesize \textbf{Key steps of Haber-Bosch reaction captured by on-the-fly MD simulation with IDEAL.} \textbf{a} The dissociation process of H\textsubscript{2} observed during the simulation. \textbf{b} The dissociation process of N\textsubscript{2} observed during the simulation. \textbf{c} The formation of NH observed during the simulation. \textbf{d} The formation of NH\textsubscript{2} observed during the simulation. \textbf{e} The formation of NH\textsubscript{3} observed during the simulation.} 
  \label{fig:example}
\end{figure}

The simulation of the entire catalytic process was carried out at 5400K on a system with a 1615-atom Iridium nanoparticle. To collect the evolution of chemical bonds, the number of bonds was monitored during the simulation as shown in \textbf{Supplementary Figure 7}. 

\begin{figure}[H] 
  \centering 
  \includegraphics[width=0.8\textwidth]{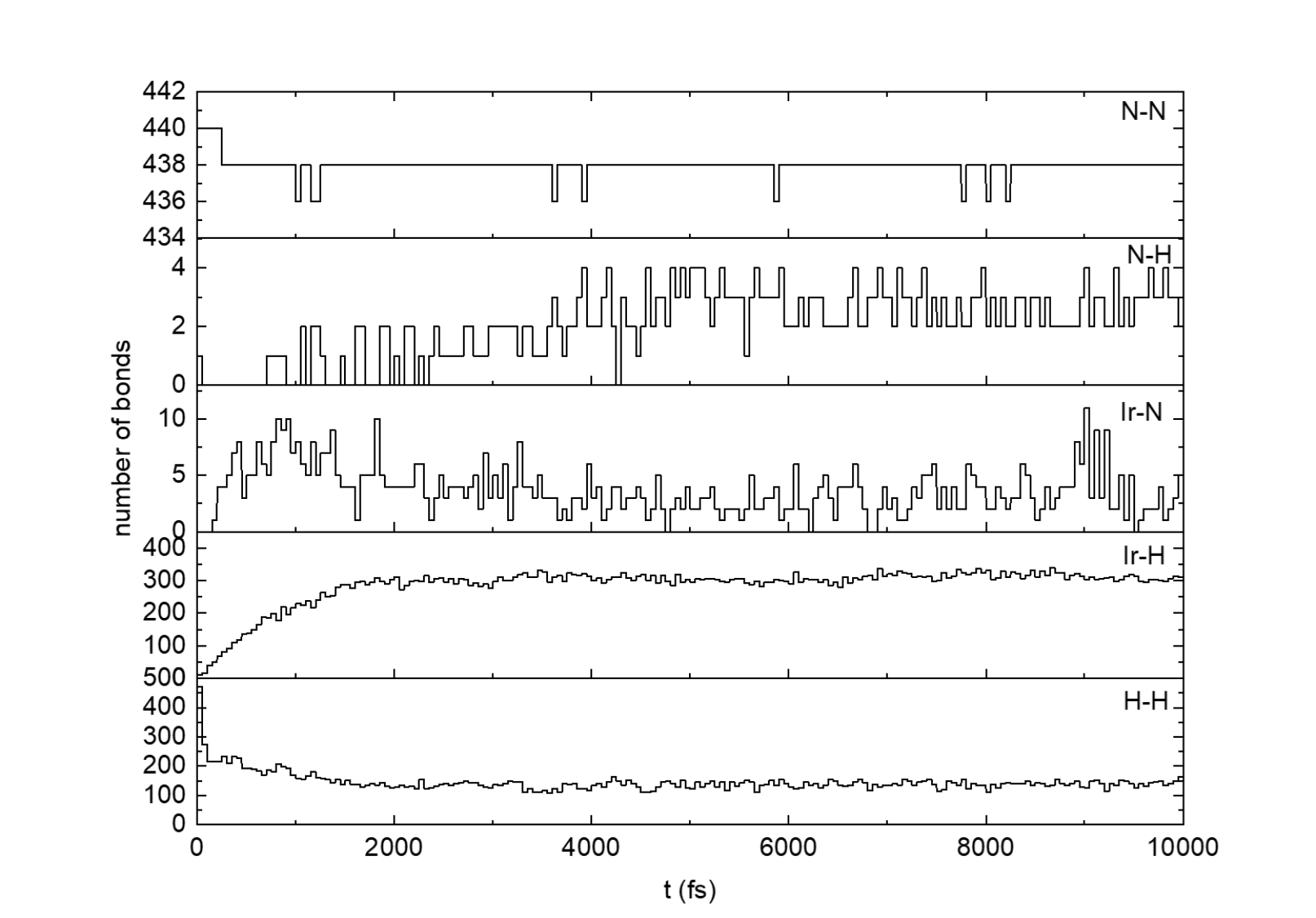}
  \caption{\textbf{The evolution of numbers of different kinds of bonds during the Haber-Bosch simulation. }}
  \label{fig:example}
\end{figure}

To estimate adsorption energies, we conducted AIMD simulations on an iridium catalytic surface. We considered isolated N, N\textsubscript{2}, H, H\textsubscript{2}, NH, NH\textsubscript{2}, and NH\textsubscript{3} structures, as well as these structures placed on the iridium catalytic surface. Based on the energies of these structures throughout the AIMD trajectories, we calculated the average energies, denoted as $\bar E(X)$, $\bar E(X_{slab})$, and $\bar E(Ir)$, where $X$ represents one of the following options: N, N\textsubscript{2}, H, H\textsubscript{2}, NH, NH\textsubscript{2}, or NH\textsubscript{3}. Here, $\bar E(X)$ stands for the average energy of the isolated $X$ structure, and $\bar E(X_{slab})$ represents the average energy of the corresponding $X$ structure placed on the iridium catalytic surface. The adsorption energy is then determined as follows:

$$
E_{abs}(X)=\bar E(X_{slab})-\bar E(X)-\bar E(Ir)
$$

The adsorption energy data obtained via DFT ($E_{abs}$) and MLFF ($\hat E_{abs}$), as well as the mean absolute errors (MAE), are presented in \textbf{Table 3} in the main context. These results demonstrate that MLFF, trained using substructure data, achieves high accuracy. For the majority of entries in the table, the deviations in adsorption energies estimated by MLFF compared to those obtained using DFT calculations are within 70 meV.


\section{Histogram plot of the distribution of the values of L}
We use $L^{(i)}$ to represent the number of substructures generated within each high-uncertainty frame during the MD simulation. We conducted on-the-fly MD simulation using the IDEAL algorithm on several lithium systems with various structure sizes and noticed that, though the upper bound for the value of $L^{(i)}$ is $N$ with $N$ representing the system size, in practice $L^{(i)}$ approaches $O(1)$ and can be treated as a constant. We show the distribution of $L^{(i)}$ value in these lithium experiments in \textbf{Supplementary Figure 8}. 

\begin{figure}[H] 
  \centering 
  \includegraphics[width=0.8\textwidth]{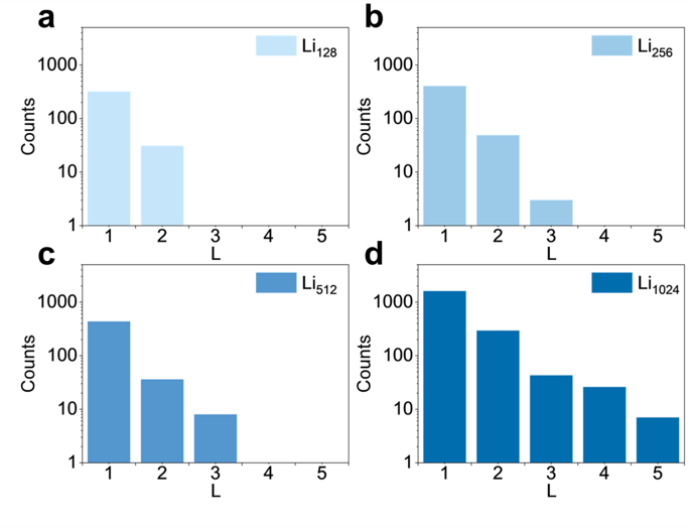}
  \caption{\textbf{The distribution of the values of L on the MD simulation of bulk-phase Li with various sizes. }} 
  \label{fig:example}
\end{figure}

\section{Molecular dynamics setups}
Molecular dynamics (MD) simulations in our experiments are all carried out using the Atomic Simulation Environment (ASE)\cite{larsen2017atomic}. In particular, we run MD under constant temperature and constant volume, i.e., NVT,\cite{frenkel2023understanding,allen2017computer} with a Langevin thermostat dynamic\cite{berendsen1984molecular,hunenberger1999ewald} simulations. For all simulations apart from those involving hydrogen, we set the time step to 1 fs, while for structures involving hydrogen, we set the time step to 0.5 fs per frame.

\section{Implementation details of the uncertainty module}
We use the SOAP descriptor\cite{bartok2013representing} in combination with the Mahalanobis distance\cite{mclachlan1999mahalanobis} for the computation of the uncertainty. In our implementation, we use the DScribe python package\cite{himanen2020dscribe, laakso2023updates} to extract the SOAP feature vector. The derivative with respect to the atomic positions is also computed using this package. To set up a SOAP descriptor, parameters as follows are needed: 

\begin{itemize}
    \item \textbf{Cutoff(Å)}: The cutoff distance (in Ångströms) specifies the radial region around each atom within which atomic positions are considered for SOAP analysis.
    \item \textbf{Max\textsubscript{l}}: Max\textsubscript{l} determines the maximum angular degree of the spherical harmonics in the SOAP analysis. 
    \item \textbf{Max\textsubscript{n}}: Max\textsubscript{n} specifies the maximum number of radial basis functions used in the SOAP analysis. 
\end{itemize}

The SOAP parameters we use for each experiment are listed in \textbf{Supplementary Table 3}. As the Mahalanobis distance is inherently differentiable with respect to the input vector, the uncertainty module possesses the ability to compute derivatives with respect to the input atomic structure.

\begin{table}[H]
\footnotesize
\centering
\begin{tabularx}{\textwidth}{XXXX} 
\toprule
{\multirow{2}{*}{\textbf{System}}} & {\multirow{2}{*}{\textbf{Cutoff (Å)}}} & {\multirow{2}{*}{\textbf{Max\textsubscript{l}}}} & {\multirow{2}{*}{\textbf{Max\textsubscript{n}}}} \\
{} &  {} & {} \\
\midrule
Li & 3.5 & 4 & 4 \\
Al\textsubscript{2}O\textsubscript{3} & 3.0 & 4 & 4 \\
Haber-Bosch & 3.5 & 4 & 4 \\
\bottomrule
\end{tabularx}
\caption{\textbf{SOAP parameters for different systems. }}
\label{tab:label}
\end{table}

In the equation introduced in \textbf{Section 4.2}, the calculation of uncertainty necessitates iterating through the known distribution $X$ to compute $\bar x$ and $\Sigma$, resulting in linear complexity with respect to the size of $X$. However, in our implementation, we employ an equivalent acceleration technique that reduces the linear complexity to O(1). To accomplish this, we introduce a vector $b$, a matrix $A$, and an index $n$ to represent the $n$-th update of the distribution $X$. The equivalent acceleration method updates these variables as $\bar x_{n+1} = \frac{1}{n+1}b_{n+1}$, $\Sigma_{n+1} = \frac{1}{n}\left[A_{n+1}-\bar x_{n+1}b_{n+1}^T-b_{n+1}x_{n+1}^T+(n+1)\bar x_{n+1}\bar x_{n+1}^T\right]
$ where $A_{n+1} = A_n+x_{n+1}x_{n+1}^T, b_{n+1} = b_n+x_{n+1}$.

\section{Quantum chemical labeling and parallelization}
Quantum chemical calculations are employed to label IDESs, thereby acquiring data for machine learning forcefields (MLFF) training. Specifically, we choose Density Functional Theory (DFT) for IDES labeling and utilize the Vienna Ab initio Simulation Package (VASP)\cite{kresse1996efficiency} to perform DFT calculations, as this method is robust and proven to be effective. The parameters we show in \textbf{Supplementary Table 4} are the typical VASP parameters for DFT calculations. DFT calculations executed using this parameter set satisfy a complexity of $O(N^3)$, where N represents the number of atoms in the structure.

\begin{table}[H]
\footnotesize
\centering
\begin{tabularx}{\textwidth}{XX}
\toprule
{\multirow{2}{*}{\textbf{Parameter}}} & {\multirow{2}{*}{\textbf{Value/Source}}} \\
{} &  {}  \\
\midrule
ALGO & Fast \\
EDIFF & 0.00135 \\
ENCUT & 520 eV \\
IBRION & -1 \\
ISIF & 3 \\
ISMEAR & -5 \\
ISPIN & 2 \\
LASPH & True \\
LCHARG & False \\
LORBIT & 11 \\
LREAL & Auto \\
LWAVE & False \\
MAGMOM & 27*0.6 \\
NELM & 100 \\
NSW & 1 \\
PREC & Accurate \\
SIGMA & 0.05 \\
KPOINTS & Automatically generated by pymatgen MPRelaxSet \\
\bottomrule
\end{tabularx}
\caption{\textbf{Typical VASP parameters for DFT calculations. }}
\label{tab:label}
\end{table}

In practical experiments, we can employ VASP parameters with some level of acceleration to expedite the research progress, accepting a tolerable degree of accuracy loss. These VASP parameters are shown in \textbf{Supplementary Table 5} and \textbf{Supplementary Table 6}. 

\begin{table}[H]
\footnotesize
\centering
\begin{tabularx}{\textwidth}{XX}
\toprule
{\multirow{2}{*}{\textbf{Parameter}}} & {\multirow{2}{*}{\textbf{Value/Source}}} \\
{} &  {}  \\
\midrule
ALGO & Fast \\
EDIFF & 0.00135 \\
ENCUT & 520 eV \\
IBRION & -1 \\
ISIF & 3 \\
ISMEAR & -5 \\
ISPIN & 2 \\
LASPH & True \\
LCHARG & False \\
LORBIT & 11 \\
LREAL & Auto \\
LWAVE & False \\
MAGMOM & 27*0.6 \\
NELM & 100 \\
NSW & 1 \\
PREC & Accurate \\
SIGMA & 0.05 \\
KPOINTS & Gamma (1,1,1) \\
\bottomrule
\end{tabularx}
\caption{\textbf{Accelerated VASP parameters for DFT calculations on Li and Al\textsubscript{2}O\textsubscript{3} systems. }}
\label{tab:label}
\end{table}

\begin{table}[H]
\footnotesize
\centering
\begin{tabularx}{\textwidth}{XX}
\toprule
{\multirow{2}{*}{\textbf{Parameter}}} & {\multirow{2}{*}{\textbf{Value/Source}}} \\
{} &  {}  \\
\midrule
ALGO & Fast \\
EDIFF & 0.01 \\
ENCUT & 600 eV \\
IBRION & -1 \\
ISIF & 3 \\
ISMEAR & 0 \\
ISPIN & 1 \\
LASPH & True \\
LCHARG & False \\
LORBIT & 11 \\
LREAL & Auto \\
LWAVE & False \\
MAGMOM & 38*0.6 \\
NELM & 100 \\
NSW & 1 \\
PREC & Normal \\
SIGMA & 0.1 \\
KPOINTS & Gamma (1,1,1) \\
\bottomrule
\end{tabularx}
\caption{\textbf{Accelerated VASP parameters for DFT calculations on Haber-Bosch systems. }}
\label{tab:label}
\end{table}

When it is needed to perform substructure embedding and DFT calculations for multiple IDESs, the process can be accelerated by distributing the substructure embedding and DFT calculations of different IDESs across multiple CPU computation nodes as illustrated in \textbf{Supplementary Figure 9}. The parallelized version of this IDEAL algorithm has also been implemented by us.

\begin{figure}[H] 
  \centering 
  \includegraphics[width=\textwidth]{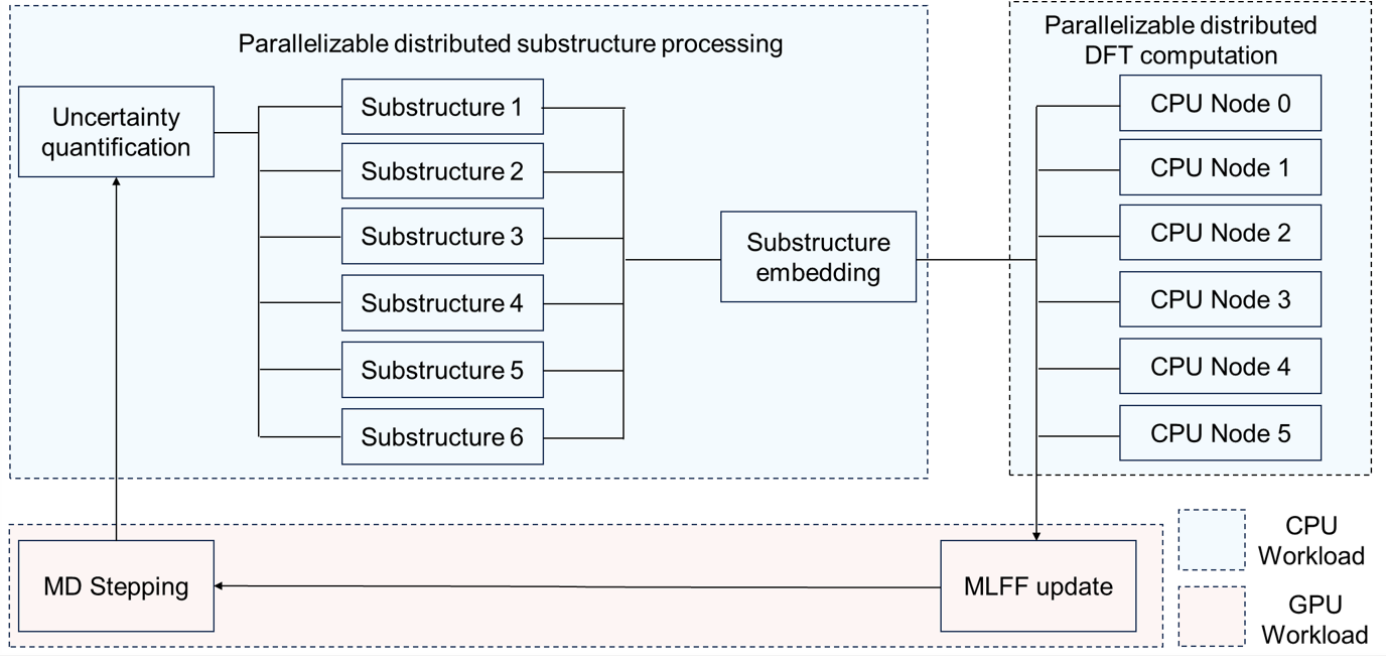}
  \caption{\textbf{A illustration of the parallel workflow of the IDEAL algorithm. } } 
  \label{fig:example}
\end{figure}

We perform a scalability check of the performance of the algorithm with respect to the number of CPU nodes. Especially, we use the melting of bulk lithium at 800K with a relatively strict uncertainty cutoff of 5. As shown in \textbf{Supplementary Figure 10}, good scalability is achieved with the increasing number of CPUs for DFT. 

\begin{figure}[H] 
  \centering 
  \includegraphics[width=0.8\textwidth]{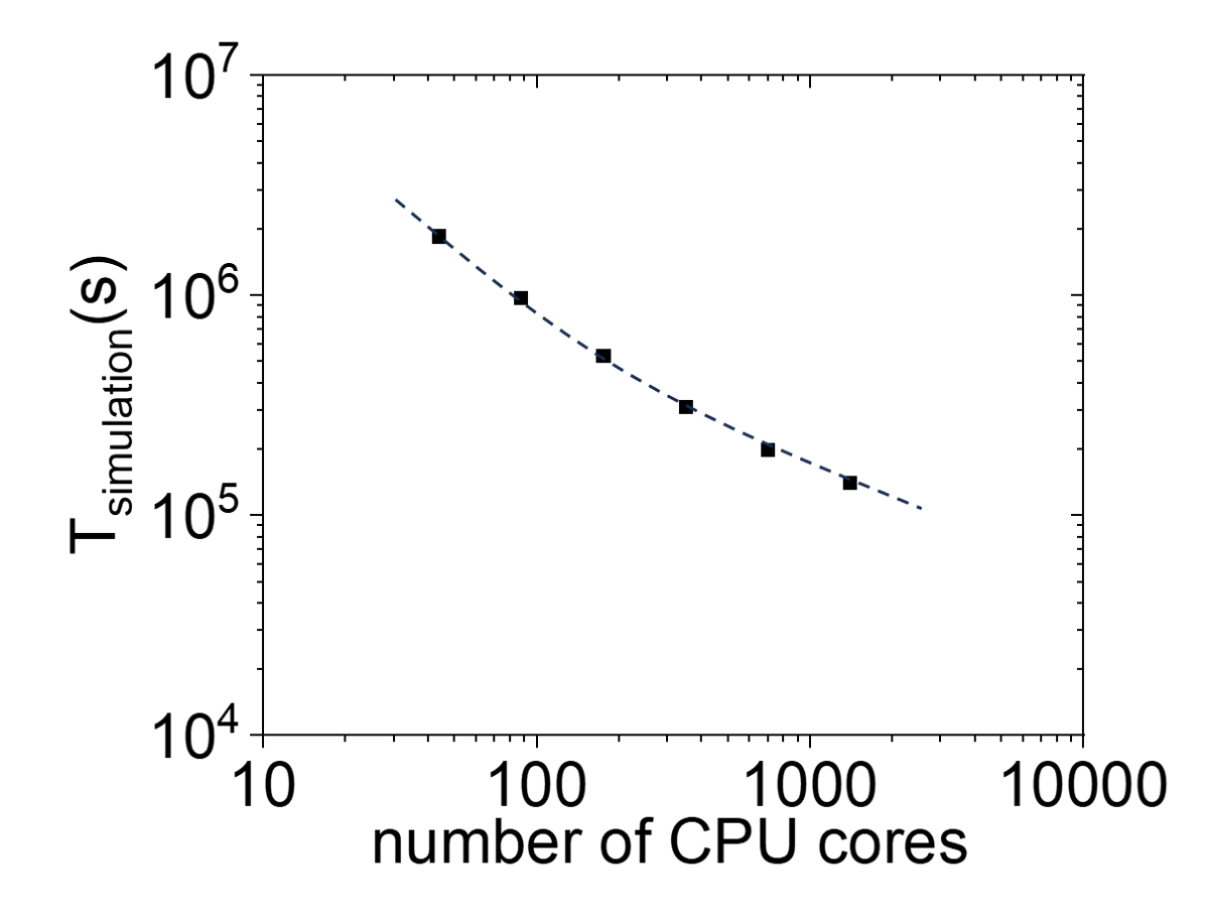}
  \caption{\textbf{Scalability of the IDEAL algorithm with respect to the number of CPU cores. } } 
  \label{fig:example}
\end{figure}

\section{Training details of machine learning forcefields}
We use a M3GNet\cite{chen2022universal} model as the main MLFF architecture. The architecture of the MLFF consists of four message-passing blocks, each containing 128 neural network units. To define the interaction range between atoms within the system, a radial cutoff of 5.0 Å is applied. For three-body interactions, a shorter radial cutoff of 4.0 Å is utilized, enabling a more localized contribution from three-body interactions to the forcefield. The values of Max\textsubscript{n} and Max\textsubscript{l}, which respectively dictate the maximum quantum number for radial basis functions and angular harmonics, are both set to 4. In the MLFF training process, we utilize a Huber loss function with a delta value of 0.1 and set the on-the-fly training learning rate to 0.0001. Detailed model parameters are listed in \textbf{Supplementary Table 7}. 

\begin{table}[H]
\footnotesize
\centering
\begin{tabularx}{\textwidth}{XXX}
\toprule
{\multirow{2}{*}{\textbf{Parameter}}} & {\multirow{2}{*}{\textbf{Value}}} & {\multirow{2}{*}{\textbf{Description}}}\\
{} & {} & {} \\
\midrule
num\_blocks & 4 & Number of message-passing blocks in the M3GNet model \\
units & 128 & Number of units or neurons in the hidden layers \\
cutoff & 5 & Cutoff for pairwise interactions \\
threebody\_cutoff & 4 & Cutoff for three-body interactions \\
Max\textsubscript{l} & 4 & Maximum quantum number for angular momentum \\
Max\textsubscript{n} & 4 & Maximum quantum number for radial basis functions \\
batch\_size & 64 & Batch size used for training \\
epochs & 50 & Number of training epochs for each update \\
lr & 0.0001 & Learning rate for model training \\
include\_forces & True & Whether forces are included in the loss function \\
include\_stresses & False & Whether stresses are included in the loss function \\
force\_loss\_ratio & 1.0 & Weighting ratio for force-related loss terms \\
stress\_loss\_ratio & 0.1 & Weighting ratio for stress-related loss terms \\
loss & Huber & Choice of loss function (Huber loss in this case) \\
delta & 0.1 & Hyperparameter delta for the Huber loss \\
\bottomrule
\end{tabularx}
\caption{\textbf{M3GNet model parameters. }}
\label{tab:label}
\end{table}

As we have mentioned before, we used a reduced-parameter MLFF on the one-million-atom lithium experiment. Here we list the detailed model parameters in \textbf{Supplementary Table 8}: 

\begin{table}[H]
\footnotesize
\centering
\begin{tabularx}{\textwidth}{XXX}
\toprule
{\multirow{2}{*}{\textbf{Parameter}}} & {\multirow{2}{*}{\textbf{Value}}} & {\multirow{2}{*}{\textbf{Description}}}\\
{} & {} & {} \\
\midrule
num\_blocks & 1 & Number of message-passing blocks in the M3GNet model \\
units & 1 & Number of units or neurons in the hidden layers \\
cutoff & 5 & Cutoff for pairwise interactions \\
threebody\_cutoff & 2 & Cutoff for three-body interactions \\
Max\textsubscript{l} & 1 & Maximum quantum number for angular momentum \\
Max\textsubscript{n} & 3 & Maximum quantum number for radial basis functions \\
batch\_size & 64 & Batch size used for training \\
epochs & 50 & Number of training epochs for each update \\
lr & 0.0001 & Learning rate for model training \\
include\_forces & True & Whether forces are included in the loss function \\
include\_stresses & False & Whether stresses are included in the loss function \\
force\_loss\_ratio & 1.0 & Weighting ratio for force-related loss terms \\
stress\_loss\_ratio & 0.1 & Weighting ratio for stress-related loss terms \\
loss & Huber & Choice of loss function (Huber loss in this case) \\
delta & 0.1 & Hyperparameter delta for the Huber loss \\
\bottomrule
\end{tabularx}
\caption{\textbf{M3GNet model reduced parameters. }}
\label{tab:label}
\end{table}

During the on-the-fly update of the MLFF, two straightforward methods exist: refining the model solely with the newly acquired data at each step or refining the model by incorporating the newly acquired data along with all previously collected data. The former approach may cause the MLFF to forget the previously accumulated data, while the latter can significantly increase the time required for MLFF updates. Here, we utilize an importance-sampling technique to enhance training efficiency, a method well-established in numerous studies for maintaining or even improving model performance while enhancing training efficiency\cite{loshchilov2015online, katharopoulos2018not}. To be specific, during each MLFF update, we refine the model using all newly added data as well as a small portion of the accumulated data. The accumulated data is sampled based on probabilities proportional to the prediction error of the MLFF on these data points: 

$$
p^{(i)}=\frac{\text{MAE}_\text{E}^{(i)}+\text{MAE}_\text{F}^{(i)}+0.1\times \text{MAE}_\text{S}^{(i)}}{\sum\limits_{j\in D}\text{MAE}_\text{E}^{(j)}+\text{MAE}_\text{F}^{(j)}+0.1\times \text{MAE}_\text{S}^{(j)}}
$$

Here, $\text{MAE}_\text{E}^{(i)}$, $\text{MAE}_\text{F}^{(i)}$, and $\text{MAE}_\text{S}^{(i)}$ represent the mean absolute error of the MLFF's predictions for energy, force, and stress, respectively, for the $i$-th structure in the accumulated dataset $D$. This method ensures that the MLFF can be updated with newly acquired data while preserving valuable information from previous data.

\bibliographystyle{unsrt}
\bibliography{supplement}